\documentclass[manuscript]{aastex631}
\usepackage{epsfig}
\usepackage{graphicx}
\usepackage{color}
\usepackage{enumerate}
\usepackage{ulem}

\def\beq{\begin{equation}}
\def\eeq{\end{equation}}

\def\pc{{\rm\,pc}}

\def\yr{{\rm\,yr}}

\def\Gyr{{\rm\,Gyr}}

\def\kms{{\rm\,km\,s^{-1}}}
\def\kpc{{\rm\,kpc}}

\def\Mpc{{\rm\,Mpc}}
\def\mbh{M_{\rm BH}}

\def\rt{r_{\rm t}}
\def\avgrt{\overline{r}_{\rm t}}
\def\avgm{\overline{m}_*}

\def\rsch{r_{\rm sch}}

\def\msun{\,{\rm M_\odot}}
\def\rsun{\rm R_\odot}
\def\mstar{\rm m_*}
\def\rstar{\rm r_*}
\def\acc{\rm M_\odot~yr^{-1}}

\def\nbody{$\textit{N}$-body~}

\def\Mbh{M_{\rm BH}}

\shorttitle{MULTI-MASS {N}-BODY MODEL}
\shortauthors{Li et al.}

\graphicspath{{./}{figures/}}

\begin{document}

\title{Tracing the Evolution of SMBHs and Stellar Objects in Galaxy Mergers: An Multi-mass Direct \nbody Model}

\author[0000-0001-6530-0424]{Shuo Li}
\affil{National Astronomical Observatories and Key Laboratory of Computational Astrophysics, \\
Chinese Academy of Sciences, 20A Datun Rd., Chaoyang District, Beijing 100012, China}
\affiliation{Department of Astronomy, School of Physics, Peking University, \\
Yiheyuan Lu 5, Haidian Qu, Beijing 100871, China}

\author[0000-0003-4121-5684]{Shiyan Zhong}
\affil{Yunnan Observatories, Chinese Academy of Sciences, \\
396 Yang-Fang-Wang, Guandu District, 650216, Kunming, Yunnan, China}

\author[0000-0003-4176-152X]{Peter Berczik}
\affiliation{Astronomisches Rechen-Institut, Zentrum f\"{u}r Astronomie, University of Heidelberg, \\
M\"{o}chhofstrasse 12-14, Heidelberg 69120, Germany}
\affil{Konkoly Observatory, Research Centre for Astronomy and
Earth Sciences, E\"otv\"os Lor\'and Research Network (ELKH),
MTA Centre of Excellence, Konkoly Thege Mikl\'os \'ut 15-17,
1121 Budapest, Hungary}
\affiliation{Main Astronomical Observatory, National Academy of Sciences of Ukraine, \\
27 Akademika Zabolotnoho St., 03680 Kyiv, Ukraine}

\author[0000-0003-2264-7203]{Rainer Spurzem}
\affil{National Astronomical Observatories and Key Laboratory of Computational Astrophysics, \\
Chinese Academy of Sciences, 20A Datun Rd., Chaoyang District, Beijing 100012, China}
\affiliation{Astronomisches Rechen-Institut, Zentrum f\"{u}r Astronomie, University of Heidelberg, \\
M\"{o}chhofstrasse 12-14, Heidelberg 69120, Germany}
\affiliation{Kavli Institute for Astronomy and Astrophysics, Peking University, \\
Yiheyuan Lu 5, Haidian Qu, Beijing 100871, China}

\author[0000-0003-3950-9317]{Xian Chen}
\affiliation{Department of Astronomy, School of Physics, Peking University, \\
Yiheyuan Lu 5, Haidian Qu, Beijing 100871, China}
\affiliation{Kavli Institute for Astronomy and Astrophysics, Peking University, \\
Yiheyuan Lu 5, Haidian Qu, Beijing 100871, China}

\author[0000-0002-5310-3084]{F.K. Liu}
\affiliation{Department of Astronomy, School of Physics, Peking University, \\
Yiheyuan Lu 5, Haidian Qu, Beijing 100871, China}
\affiliation{Kavli Institute for Astronomy and Astrophysics, Peking University, \\
Yiheyuan Lu 5, Haidian Qu, Beijing 100871, China}

\begin{abstract}

By using direct \nbody numerical simulations, we model the dynamical co-evolution of two supermassive black holes (SMBHs) and the surrounding stars in merging galaxies. In order to investigate how different stellar components evolve during the merger, we generate evolved stellar distributions with an initial mass function. Special schemes have also been developed to deal with some rare but interesting events, such as tidal disruption of main sequence stars, the plunge of low mass stars, white dwarfs, neutron stars and stellar mass black holes, and the partial tidal disruption of red giants or asymptotic giant branch stars. Our results indicate that the formation of a bound supermassive black hole binary (SMBHB) will enhance the capture rates of stellar objects by the SMBHs. Compared to the equal stellar mass model, the multi-mass model tends to result in a higher average mass of disrupted stars. Instead of being tidally disrupted by the SMBH, roughly half of the captured main sequence stars will directly plunge into the SMBH because of their small stellar radius. Giant stars, on the other hand, can be stripped of their envelopes if they are close enough to the SMBH. Though most remnants of the giant stars can survive after the disruption, a small fraction still could plunge into the SMBH quickly or after many orbital periods. Our results also indicate significant mass segregation of compact stars at the beginning of the merger, and then this effect is destroyed as the two SMBHs form a bound binary.

\end{abstract}


\section{Introduction} \label{intro}

Supermassive black hole binaries (SMBHBs) are predicted as the descendents of the hierarchical galaxy formation model \citep{bege80,volo03}. Over the past few decades, more and more observational evidence has indicated that most massive galaxies, if not all, have a supermassive black hole (SMBH) hidden in the center \citep{korm13}. Since massive galaxies could undergo several mergers in their evolutionary history, it is natural to predict the existence of SMBHBs in merging or merged galaxies. Besides, many investigations find that there is a close connection between the SMBH and its host galaxy \citep{mago98,ferr00,gebh00,trem02,korm13}. Therefore it is essential to find out how SMBHBs evolve in galaxy mergers.

In a merging galaxy, two SMBHs will first approach each other mainly through dynamical friction. However, as two SMBHs get closer, this effect gets weaker and weaker. When two SMBHs are close enough to form a bound binary system, the dynamical friction is not efficient enough to drive the SMBHB coalesce within Hubble time in most of the massive galaxies \citep{bege80}. A hard SMBHB can eject surrounding stars to transfer their orbital energy and angular momentum, which may be efficient to drive two SMBHs coalesce quickly \citep{sasl74,mikk92,quin96}. But it needs enough close encounter stars to be ejected by the SMBHB, which may be not always the case, because stars scattered into the vicinity of the SMBHB through two-body relaxation may be inefficient in spherical stellar distributions \citep{bege80,yu02,milo03,ber05}. Fortunately, both gas dynamics and more realistic stellar dynamics other than spherical two-body relaxation can avoid this problem \citep{goul00,chat03,mer04,ber06,pret11,khan11}.

There are many observational evidence to confirm the above scenario, such as dual active galactic nuclei (AGNs), jet reorientation in X-shaped radio galaxies, double-peaked emission lines, quasi-periodic outbursts in some blazars and so on \citep{liu02,komo03,liu04,shen13,komo20,tang21}. Most of these are indirect evidence of SMBHBs and they are in gaseous environments. In gas poor environments, hard SMBHBs are very difficult to detect. Besides emitting electromagnetic waves, a close SMBHB has strong gravitational wave (GW) emission, which could be detected by the ongoing Pulsar Timing Arrays (PTAs) and the planned space borne GW detectors such as the Laser Interferometer Space Antenna (LISA)\footnote{https://www.elisascience.org/} , Taiji Program and Tianqin Project\footnote{http://tianqin.sysu.edu.cn/en/} \citep{fost90, verb16, mei21, luo21}. However, after years of effort, though PTAs may already get some clue on the stochastic GW background, the specific detection is still missing \citep{shan15,baba16,arzo16,arzo20}. There are also other processes connected to SMBHs and SMBHBs, which could generate GW emission for LISA or similar instruments. For instance, compact stellar objects can inspiral to SMBHs with detectable GW emissions for LISA, which is the so-called extreme mass ratio inspirals (EMRIs) \citep[][and references therein]{pau18}. Similar events should also exist in the SMBHB system. But it has not been well studied. Nevertheless, due to limited spatial resolutions of GW detectors, even if a GW signal from the SMBHB could be detected, it is still difficult to locate its host galaxy. The inconsistency between theoretical expectation on SMBHBs and GW observations may be due to many processes that have not been clarified. It is important to design observations that can be used to best constrain the evolution of SMBHBs.

A dormant SMBH can be temporarily illuminated by tidal disruption events (TDEs). If a star closely approaches the SMBH by less than a critical distance, it will be torn into debris by the SMBH, which may prompt temporary flares with periods from days to years \citep{hil75,ree88,evan89,guil13}. We use for the critical distance  tidal radius
\beq
\rt\backsimeq \mu\rstar(\Mbh/\mstar)^{1/3},
\label{eq:rt}
\eeq
where $\mu$ is a dimensionless parameter of order unity which reflects the stellar structure, $\rstar$, $\mstar$ and $\Mbh$ are the stellar radius, the stellar mass and the mass of black hole (BH), respectively. Since the first identified event in the 1990s, dozens of TDEs have been reported, with emission range from $\gamma$-ray to radio bands \citep[][and references therein]{komo99,komo15,geza21}. Similar events can also happen in SMBHB systems, but light curves may be different from normal single SMBH TDEs. Due to the perturbation of the companion SMBH, a TDE in SMBHB system could have repeated gaps or be truncated, which has been investigated theoretically and numerically \citep{liu09,rica16,coug17,vign18}. With the expansion of the TDE sample in these years, a few SMBHB TDE candidates also have been founded in observation \citep{liu14,shu20,huan21}. With more and more optical and X-ray transient surveys, such as the Large Synoptic Survey Telescope (LSST)\footnote{https://www.lsst.org}, the All-Sky Automated Survey for Supernovae (ASAS-SN)\footnote{http://www.astronomy.ohio-state.edu/asassn/index.shtml}, and the Einstein Probe (EP)\footnote{http://ep.bao.ac.cn/}, join the game, the sample of the SMBHB TDE could significantly increase in the near future.

There are some attempts to investigate the tidal disruption rate (TDR) of SMBHBs. Theoretical analyses and numerical scattering experiments indicate that, due to the perturbation of the companion SMBH and the triaxial stellar distribution, TDR in galaxy merger remnants could be enhanced from a few times to a few orders of magnitudes \citep{ivan05,chen09,wegg11,liu13}. However, since the interaction of two SMBHs and surrounding stars is chaotic during the formation of the SMBHB, these results are limited and can not fully reveal the underlying physical processes. In order to investigate the dynamical co-evolution of SMBHBs and stars in galaxy mergers with more depth, we used GPU accelerated direct $\nbody$ simulations to analyze a series of models on the TDR evolution of SMBHBs in both major mergers \citet[][hereafter Paper I]{li17} and minor mergers \citet[][hereafter Paper II]{li19}. Both equal and unequal mass models indicate significantly enhanced TDRs during two SMBHs forming bound binary systems, which can be considered as a possible explanation for the high detection rates of TDEs preferred in E+A galaxies \citep{arca14}. However, for simplicity, it was assumed in both papers that all stars are solar type stars with the same mass. In reality, for example, giant stars will be partially tidally disrupted by the SMBH at relatively large distances. Compact objects (neutron stars, black holes), on the other hand, could directly plunge into the SMBH without disruption. Even main sequence stars could have different fates after considering different masses. According to Eq.~\ref{eq:rt}, massive main sequence stars with larger stellar radius correspond to larger tidal radius. Low mass main sequence stars usually correspond to smaller tidal radius. Massive stars with large envelopes may as well have only partial mass loss from TDE (see e.g.\cite{zhong22}). Finally, instead of a tidal disruption with significant flare, many main sequence stars near the low mass end actually will be entirely swallowed by the SMBH, because their tidal radii are very close to the Schwarzschild radius of the SMBH.

In this work we improve our model by using a realistic distribution of stellar properties (mass and radius) instead of equal mass stars as in previous papers such as e.g. \citet{li19}. In a dry major merger case, the typical evolution time is $\sim 1\Gyr$ \citep{colp14}. This period of time is long enough for a stellar system to evolve from the zero-age main sequence to different stellar components. For this reason, our stellar population is initialized with an age of  $\sim 1\Gyr$ to account for the lifetime of the galaxies before their merger. We developed a special scheme to deal with TDEs of different types of stars, which includes variation of the tidal disruption radius as a function of stellar parameters, partial tidal disruption of giant stars, and direct plunges of compact objects (see for details next section and Fig.~\ref{fig:Scheme}). With these modifications we are able to trace the co-evolution of all kinds of stars and SMBHs in galaxy mergers. This paper is organized as follows. We introduce our simulation models in Section~\ref{mod}. Our simulation results on different types of stars are presented in Section~\ref{res}. In Section~\ref{Diss} we discuss how to extrapolate our results to more realistic systems, and some observational implications are also provided. A short summary is allocated to Section~\ref{sum}.

\section{The Direct \textit{N}-body Model With More Realistic Stellar Objects}
\label{mod}

Our numerical model base on a direct \nbody code $\varphi$\,-{\sc Grape}/$\varphi$\,-GPU, which is an accurate GPU accelerated code with fourth-order Hermite integrator and efficient parallel scheme \citep{maki92,ber05,harf07}. $\varphi$\,-{\sc Grape}/$\varphi$\,-GPU has been proved to be an efficient tool on investigating the dynamical co-evolution of SMBH and stars in both single galaxy and galaxy mergers \citep{ber06,gual08,khan11,pret11}. It can be also adapted to investigate the tidal disruption evolution, with a simplified tidal disruption scheme included \citep{zhong14,li17,pana18}. Recently, \citet{khan18} have tried to study the coalescence time of supermassive black holes in galaxy mergers with post-Newtonian (PN) terms and stellar mass function included. In this work, we try to introduce a more realistic model, which can self-consistently study how different type stellar objects, such as main sequence(MS) star, red giant (RG), asymptotic giant branch(AGB), white dwarf (WD), neutron star (NS), stellar mass black hole (BH), interact with SMBHs in galaxy mergers.

\subsection{Initial Mass Function and Stellar Evolution}
\label{sse}

Compared with previous equal stellar mass models, a more realistic multi-mass model with initial mass function is adopted. We assume that the stellar mass of a star cluster ranges from $0.1\msun$ to $100\msun$, and the initial mass function follows a multiple-part power-law as \citet{krou01} suggested. This model corresponds to an average initial mass $\overline{m}_*\sim0.6 \msun$.

For simplicity, we are focusing on dry mergers with gas poor environments, which are usually dominated by early type galaxies, and the starbursts induced by mergers are not significant. It is reasonable to assume that all the stars have evolved for a relatively long time as the initial condition. We evolve all the stars for $1~\Gyr$ by using the stellar evolution package SSE \citep{hurl00}. After the evolution, our model with largest particle number $N = 10^6$ has $4484$ giant branch stars (includes RGs, core He burning stars, early AGBs and thermally pulsing AGBs), $29811$ WDs(includes C/O WDs and O/Ne WDs), $5246$ NSs and 1884 BHs. And the average mass decreased to $\overline{m}_*\sim0.4 \msun$ due to the mass loss during the stellar evolution. The same scheme as \citet{pana19} adopted is involved to model the natal kick during the formation of NSs and BHs. The kick amplitude of NSs is represented by a Maxwellian distribution with 1D velocity dispersion $\sigma = 265 \kms$ \citep{hobb05}. For BHs, both the mass and kick velocity sensitively depend on the "fallback" of debris, those materials failed to escape during the explosion of the progenitors \citep{colg71,chev89,zhan08}. We take this effect into account by following the scheme suggested by \citet{belc02}.

As a result, a group of stars with different stellar components including NSs and BHs with initial kick velocities have been generated. In order to model a dense star cluster, the next step is to assign position and velocity to every star according to a proper stellar mass distribution.

\subsection{Dense Star Cluster Model}
\label{distr}

The dynamical parameters of the initial condition are similar to those in Paper I. The stellar distribution of the dense star cluster around the SMBH is represented by a Dehnen model \citep{deh93}.

\beq \rho(r)=\frac{3-\gamma}{4\pi}\frac{Ma_{\rm D}}{r^\gamma(r+a_{\rm D})^{4-\gamma}} ,
\label{eq:Dehnen rho}
\eeq
where $a_{\rm D}$, $M$ and $\gamma$ denote the scaling radius, the total mass of the galaxy/nucleus and the density profile index, respectively. Here we adopt the units ${\rm G}=M=a_{\rm D}=1$, and assume $\gamma=1$ in the following discussion for simplicity, because models with more steep central density profiles are very time consuming in the integration. The influence of $\gamma$ has been carefully discussed in Paper I, with equal stellar mass models. The general results should be still informative here. According to results in Paper I, steep density profiles usually correspond to significantly higher TDRs, and the boosted TDRs in phase II are common for all models with different $\gamma$. However, the magnitude of the enhancement of the averaged TDRs in phase II only weakly depends on $\gamma$. Detailed discussions can be found in Paper I. The relation between numerical and physical quantities can be derived as

\begin{eqnarray}
[\rm{T}] &=& \left(\frac{{\rm G}M}{a_{\rm D}^3}\right)^{-1/2} \nonumber \\
    &=& 1.491\times 10^7(2^{\frac{1}{3-\gamma}}-1)^{3/2}\left( \frac{M}{10^{9}\msun}\right)^{-1/2}\left(\frac{r_{1/2}}{1\kpc}\right)^{3/2} \yr, \\
    \label{eq:scalingT}
[\rm{V}] &=& \left(\frac{{\rm G}M}{a_{\rm D}}\right)^{1/2} \nonumber \\
    &=& 65.58\times (2^{\frac{1}{3-\gamma}}-1)^{-1/2}\left( \frac{M}{10^9\msun}\right)^{1/2}\left(\frac{r_{1/2}}{1\kpc}\right)^{-1/2} \kms, \\
    \label{eq:scalingV}
[\rm{R}] &=& a_{\rm D}=(2^{\frac{1}{3-\gamma}}-1)\left(\frac{r_{1/2}}{1\kpc}\right) \kpc,
    \label{eq:scalingr}
\end{eqnarray}
\begin{eqnarray}
[\dot{\rm{M}}] &=& M / [\rm T] \nonumber \\
          &=& 67.07\times(2^{\frac{1}{3-\gamma}}-1)^{-3/2}\left( \frac{M}{10^9\msun}\right)^{3/2}\left(\frac{r_{1/2}}{1\kpc}\right)^{-3/2} \msun/\yr.
\label{eq:scalingMdot}
\end{eqnarray}

The SMBH is represented by a heavy particle with mass $\mbh = 0.01$ at the center. The Dehnen model above does not considered the influence of SMBH and multiple stellar mass distribution. Similar to Paper I, to relax the SMBH with surrounding stars, we integrate the entire system for dozens of \nbody unit time before the simulation, which is roughly the two-body relaxation timescale at the influence radius of the SMBH in the model. After this relaxation procedure, the mass segregation of heavy stellar components in the central region is significant. It has been confirmed by the result in the left panel of Fig.~\ref{fig:LagrCpt} Based on this template model, two identical galaxies/nuclei are set in a parabolic orbit with initial separation $d \sim 20$ and the first pericenter $\sim 1$. The pericenter distance here is for the convenience of comparison with Paper I. A closer encounter will lead to faster evolution, but the general results should be similar. All the integrations are executed on the $laohu$ GPU cluster in National Astronomical Observatories of China (NAOC).

\subsection{Scheme of Close Encounters with SMBHs}
\label{tdschm}

The interactions between merging SMBHs and the surrounding stars, especially those stars close to SMBHs, are the focus of this work. In order to carefully investigate these "close encounters" we involve some special treatments. In general, there are four kinds of close encounters in our model: normal stars tidally disrupted by the SMBH, main sequence stars with relatively light mass swallowed by the SMBH without tidal flares, giant branch stars partially disrupted, and compact stars, such as NSs and BHs, forming EMRIs or the extreme mass ratio bursts(EMRBs, more discussions can be found in Seciton~\ref{gw}).

\begin{figure}
\begin{center}
\includegraphics[width=0.8\textwidth,angle=0.]{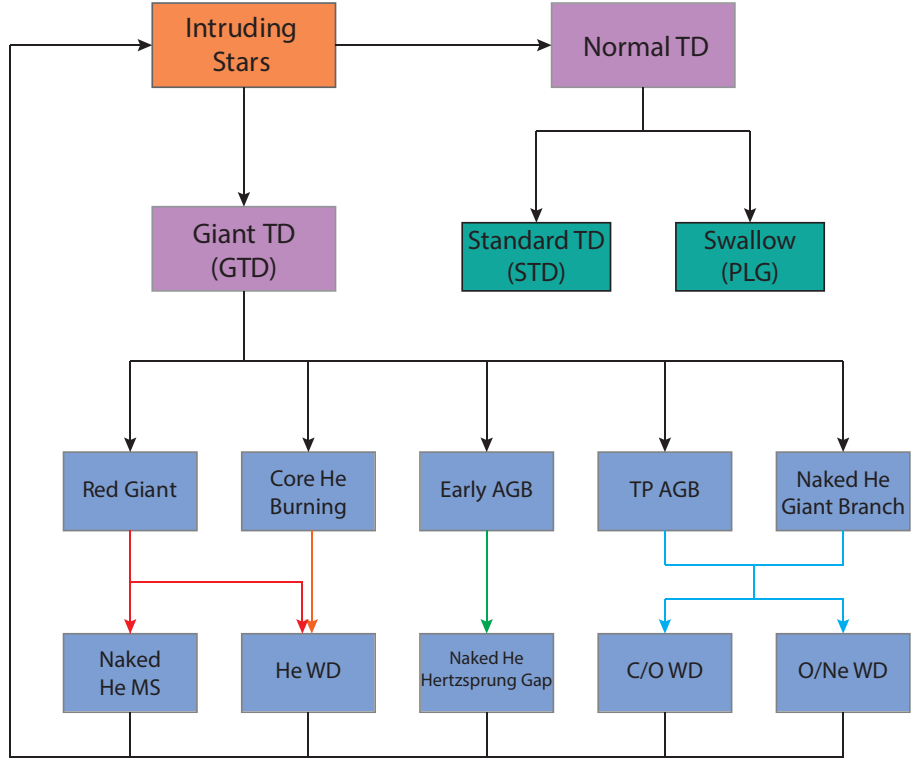}
\caption {The fate of different types of stars after close encounters with SMBHs.
\label{fig:Scheme}}
\end{center}
\end{figure}

Fig.~\ref{fig:Scheme} demonstrates what will happen if an ordinary star passes by a SMBH within the tidal radius. Most of stars in our model are MS stars. If the stellar mass of a star is massive enough, there will be a typical TDE. Conversely, the tidal radius of a relatively light MS star or WD could be smaller than the Schwarzschild radius. The star will be directly swallowed by the SMBH without flare. The critical swallow radius of a BH, comparing to the Schwarzschild radius $\rsch$, can be enlarged when we take into account the eccentricity of the intruder star. In this work we adopt the pericenter of the marginally stable orbit $r_{\rm MSO}$ as the critical boundary. A star will be marked as a plunge event when it's tidal radius and separation to the SMBH are both smaller than $r_{\rm MSO}$ \citep{cutl94, gair05}

\beq
r_{\rm MSO} = \frac{3+e}{1+e} \rsch,
\label{eq:MSO}
\eeq
where $\rsch=2\rm{G}\mbh/\rm{c}^2$ is the Schwarzschild radius, $e$ is the eccentricity of the star. $r_{\rm MSO}$ determines the minimum pericenter distance of a test particle with fixed eccentricity. A test particle with pericenter distance smaller than $r_{\rm MSO}$ plunges directly into the black hole. If $e=0$ the $r_{\rm MSO}$ is equivalent to the innermost stable circular orbit(ISCO). For extreme hyperbolic situations with $e\rightarrow \infty$, only orbits with pericenter distances larger than the $\rsch$ can survive. For simplicity, the mass and linear momentum of both the disrupted and swallowed stars will directly add to the SMBH in our model. Here we only take into account the linear momentum, and the angular momentum is not considered because we can not trace the spin evolution of the SMBH without the PN approximations included.

In addition to main sequence stars, there are many giant branch stars (GBs), such as RGs and AGBs. These stars, with their compact cores and diffuse envelopes, usually have very large radii compared with MS stars, which consequently corresponds to significantly large tidal radii. In most cases a giant tidal disruption (GTD) will not cause a complete disruption. Inversely, the stellar core, and even a large fraction of the envelope of the star can survive \citep{macl12}. We assume that the envelope will be completely striped and accreted by the SMBH immediately after the GTD, and the remnant could be considered as a production of a fast evolved giant star. According to the predicted evolution paths proposed by \citet{hurl00}, we make a simplified evolution scheme of GTDs in Fig.~\ref{fig:Scheme}. Depending on the initial mass of the disrupted star, GTDs in this scheme may result in different types of remnants, such as naked helium main sequence stars, WDs and naked helium stars in the Hertzsprung gap. More details can be found in the discussion of \citet{hurl00}.

Compact stars, on the other hand, are simply assumed to be directly swallowed by the SMBH if their separations to the SMBH are smaller than the ISCO. In that case, the linear momentum and mass of the compact stars will be added to the SMBH. Therefore the capture criteria of compact stars is different from other stars. Our integration does not include the post-Newtonian approximation because it is very difficult to be involved in three body problems. The orbits of those compact stars with heavier masses, especially when they are close to ISCO, are not accurate enough. The eccentricity evolution due to GW emission can not be well traced in this work. We prefer to leave this problem in future works. For this reason we did not adopt the $r_{\rm MSO}$ as the criteria for simplicity.

\subsection{From numerical models to the reality}
\label{scl}

Direct \nbody simulations are limited by particle resolution. Even taking into account the acceleration of Heterogeneous Computing, the maximum particle number that can be managed by direct \nbody simulation with reasonable integration time is only several million, which is obviously less than the number of stars in a typical galaxy. A direct consequence of limited particle resolution is that the TDEs are so rare in the simulation that we can not collect enough events to make statistical analyses. For example, if there is a galaxy with total mass $10^9 \msun$, and the half mass radius is $r_{1/2} = 1 \kpc$, then a length scaling relation can be set up by Eq.~\ref{eq:scalingr}. For a solar type star, the tidal radius is $\sim 10^{-5} \pc$ corresponding to $\sim 10^{-8}$ in this simulation unit, which is a very tiny scale in the simulation. The collected TDEs in such configuration will be only a few in the entire simulation. In order to collect more TDEs in the simulation, we have to adopt a larger "tidal radius", which is $\rt\sim 10^{-4}$. And the Schwarzschild radius also adopts the same scaling to make sure that $\rt/\rsch$ is the same both in simulation and reality. By integrating several models with different tidal radii, we can try to extrapolate the simulation result to reality. This scheme has been proved to be feasible in Paper I. Since $\rt\propto\Mbh^{1/3}$, while $r_{\rm MSO}\propto\Mbh$, it is not recommended to make direct extrapolations from simulation results to other real systems with different $\Mbh$. Otherwise the results of the normal TDEs and those swallow events could be inconsistent. For this reason, we have to fix the mass of the SMBH in our simulations. In the following discussion, we set the total mass of each galaxy/nucleus is $10^9\msun$, and the mass of each SMBH is $10^7\msun$.

The limited particle resolution also induced other artificial effects. Some dynamical processes are $N$ dependent. For instance, the two-body relaxation timescale follows $t_{\rm r}\propto N/\ln N$, which means the two-body relaxation in our model will be much faster than the case of a real galaxy. In a spherical stellar system with a SMBH in the center, two-body relaxation is the most important mechanism to scatter stars into the tidal disruption loss-cone, the phase space region corresponding to the orbits of stars can be tidally disrupted \citep{fra76,lig77}. The process that scatters stars into loss-cone is the so-called loss-cone refilling. The situation in merging galaxies is quite different. In principle, as we did in Paper I, we can divide the dynamical evolution of the SMBHB in a galaxy merger into three phase. In phase I, two galaxies and their central SMBHs have so large separation that the interaction between central SMBH and surrounding stars is roughly the same as the case of a single SMBH in an isolated galaxy. That means the loss-cone refilling will be dominated by two-body relaxation. Thus the TDEs rate will be $N$ dependent in phase I. As two SMBHs get more and more close to each other, the perturbation becomes stronger and stronger in phase II. As a result, the perturbation will dominate the loss-cone refilling, which will result in an approximate $N$ independent TDEs rate. In phase III, two SMBHs form a compact binary and the loss-cone refilling is very complicated. By carefully analyzing the different situations in different phases, it is possible to make a credible extrapolation based on analytical models and our numerical models with different $N$. Detailed discussions can be found in Paper I and II. There are more discussions in Section~\ref{extrap}.

In addition to the artificial effects mentioned above, the stellar evolution model also needs to consider the influence of the limited particle resolution. For instance, if we use $10^6$ particles to represent a dense star cluster, the total mass of stars in the Kroupa model should be $\sim 6\times10^5 \msun$. However, we have to adapt this cluster to represent a galaxy with $10^9$ stars. That means every particle in our model actually corresponds to a group of $1000$ stars in the real galaxy. A straightforward discrepancy between our model and the real galaxy is that they have different two-body relaxation timescale. For example, we can assume that all the stars in the cluster have evolved for $1\Gyr$. In our model, due to the limited particle number, the system could be well relaxed. While the same thing should not happen in a real dense star cluster with $10^9$ stars. Therefore we need to have a proper scaling relation between the stellar evolution timescale in reality and the two-body relaxation timescale in our model. A simplified solution is to make sure that the ratio between relaxation timescale and stellar evolution time scale in the real dense star cluster should be equal to the ratio in our model \citep{pana19}. This scaling relation is essential for models with stellar evolution included (more discussions can be found in \citet{pana19}). However, as we mentioned before, we are focusing on the merging phase, which corresponds to a relatively short period of time. Consequently, in dry mergers with all stars that have already evolved for quite a long time, the stellar evolution during the merger could be neglected. Instead of including stellar evolution, it is more important to have an initial model with properly evolved stars before the merger, which is adopted here.

\section{Results}
\label{res}

\subsection{Dynamical evolution of SMBHs and surrounding stars}
\label{devo}
\begin{figure}
\begin{center}
\includegraphics[width=0.8\textwidth,angle=0.]{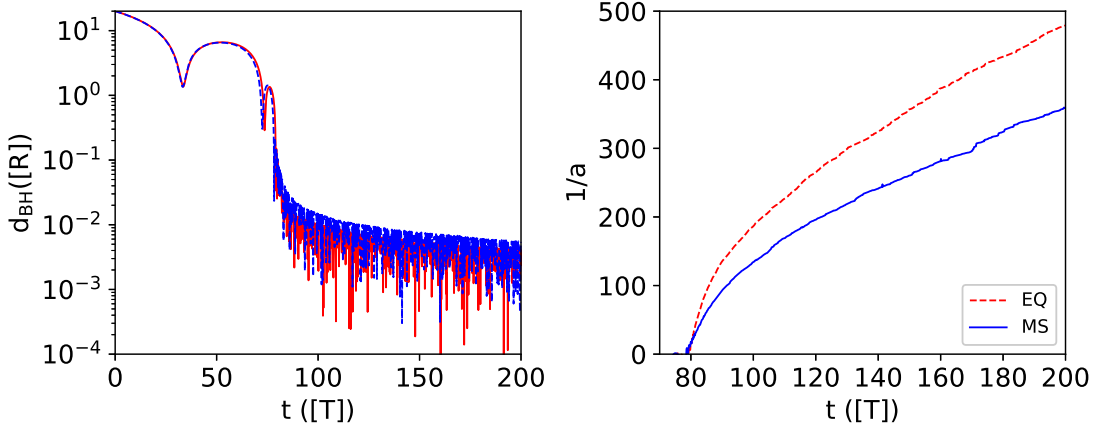}
\caption {The dynamical evolution of two SMBHs in merging galaxies. Red dashed and blue solid lines correspond to equal stellar mass model and multi-mass model, respectively. The left panel represents the separation evolution of two SMBHs. The right panel represents the semi-major axis evolution of the SMBHB, after a bound binary system is formed.
\label{fig:para}}
\end{center}
\end{figure}

Compared with previous major merger models with equal mass stars, we have more realistic models with different stellar components. Fig.~\ref{fig:para} demonstrates the dynamical evolution of SMBHs in merger galaxies, which are numerically integrated by our models with $N=2\times10^6$ particles. Unless otherwise specified, all results in following discussions are based on this model. For better comparison, we have two integrations with and without initial mass function (IMF), which are represented by blue solid lines and red dashed lines, respectively. The left panel represents the evolution of $d_{\rm{BH}}$, the separation of two SMBHs, and the right panel demonstrates the evolution of the semi-major axis $a$. Since bound SMBHBs are not formed in phase I, the time begins from $t\sim 80$ in the figure. As the results indicated, though the evolutions of two SMBHs in different models are roughly the same in phase I, there is a significant difference in phase II. The SMBHB in the model with IMF evolves slower than the case of the equal mass model, which looks inconsistent with \citet{khan18}. However, our integrations only continued to $\sim 100$ \nbody time unit, which is far less than \citet{khan18} did. According to the Fig.~1 and Fig.~2 in their paper, the difference between equal mass and multi-mass models is not obvious in the early stages of the SMBHB formation.

\begin{figure}
\begin{center}
\includegraphics[width=0.8\textwidth,angle=0.]{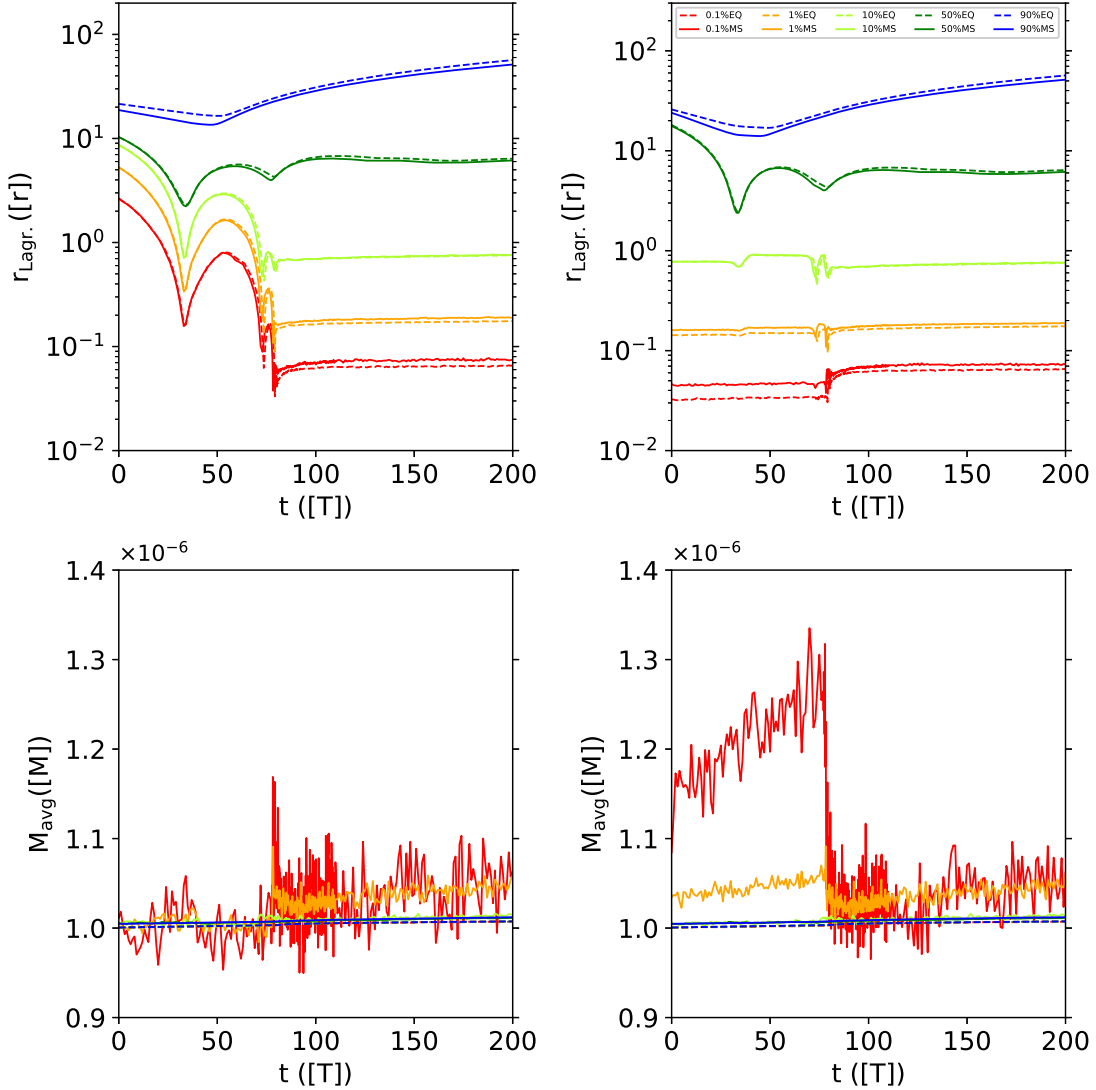}
\caption {The evolution of Lagrangian radii and corresponding average mass. The upper left panel is the Lagrangian radii relative to the center of mass, and the upper right panel corresponds to one of the SMBH. Dashed lines and solid lines denote equal mass and multi-mass models respectively. The bottom left and right panels demonstrate corresponding average mass inside each Lagrangian radius. Different colors represent different mass fractions of the total mass.
\label{fig:LagrComp}}
\end{center}
\end{figure}

The evolution of Lagrangian radii is demonstrated in Fig.~\ref{fig:LagrComp}. A Lagrangian radius is a sphere with the center of the stellar system. Stars inside this sphere occupy a fixed fraction of the total mass of the system. By tracing the evolution of several Lagrangian radii with different mass fractions, we can trace the dynamical evolution of the star cluster. Here dashed lines and solid lines denote equal mass and multi-mass models respectively. Different colors correspond to $0.1\%$, $1\%$, $10\%$, $50\%$, and $90\%$ of the total mass respectively, which have been marked in the legend. It should be noticed that the $1\%$ curve is close to the influence radius, because the stellar mass inside this radius is close to the mass of the central SMBH. The upper left panel of the figure represents the Lagrangian radii relative to the center of mass. Since two galaxies are far away from each other, there are only a few stars around the center of mass, which results in the large Lagrangian radii at the beginning. The upper right panel demonstrates the Lagrangian radii relative to one of the SMBH. The inner region around the SMBH of the equal mass model is slightly more compact than the multi-mass model, which is consistent with a similar multi-mass model with single BH \citep{bau04}. The bottom panels are the corresponding average mass inside each Lagrangian radii. In the equal mass model, all average masses are equal to $10^{-6}$, the single particle mass in the simulation. In the multi-mass model, the average masses inside larger radii, which correspond to all Lagrangian radii equal and larger than $10\%$ in the figure, are very close to $10^{-6}$. However, this is not the case in the inner region. According to bottom panels, the average mass inside Lagrangian radii of mass fraction $0.1\%$ and $1\%$ are significantly heavier than the average mass of the entire system, which can be considered as the consequence of the mass segregation at the inner region. In the bottom left panel, the average masses of the inner region are close to $10^{-6}$ at the beginning of the integration, because the center of mass is at the midpoint where the stellar density is very low. The mass segregation effect is more significant in the plot relative to one SMBH, which is demonstrated in the bottom right panel. The inner region average masses are significantly higher than the equal mass model at the beginning of the integration. However, during the formation of the bound SMBHB, the central average masses decline sharply. The strong interaction between the two SMBHs and the violent evolution of stars in the central region suppress the mass segregation quickly. After the two SMBHs form a compact binary, the average mass at the central region starts to recover with gradually increasing.

\begin{figure}
\begin{center}
\includegraphics[width=0.8\textwidth,angle=0.]{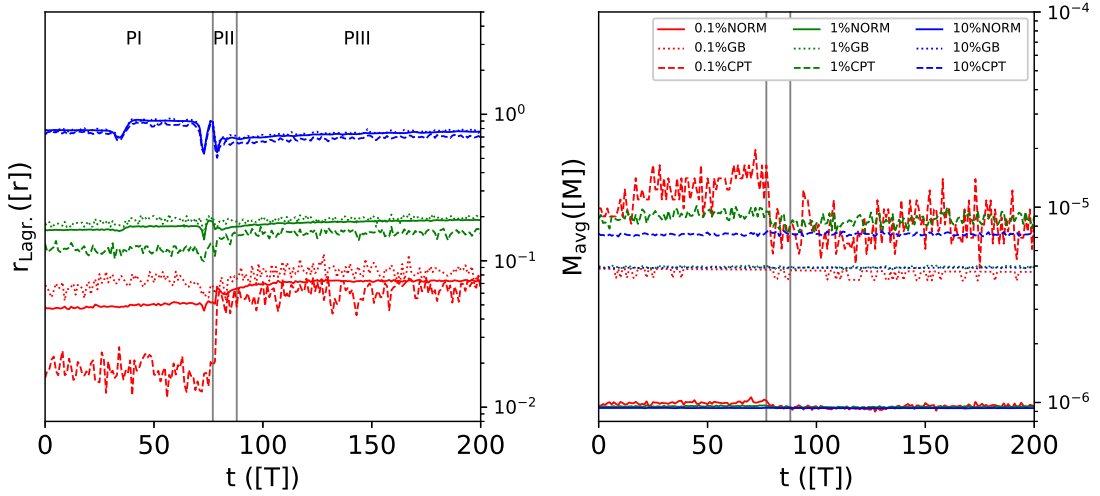}
\caption {The evolution of Lagrangian radii and corresponding average masses of different components relative to one of the SMBH. Solid lines, dotted lines and dashed lines denote normal stars, GBs, and compact stars, respectively. Different colors represent different mass fractions of the total mass of each component. The entire evolution has been divided into three phases by two vertical gray lines. The left panel is the evolution of Lagrangian radii, and the right panel is the evolution of average mass inside each Lagrangian radius.
\label{fig:LagrCpt}}
\end{center}
\end{figure}

Fig.~\ref{fig:LagrCpt} demonstrates the evolution of Lagrangian radii and corresponding average masses of different components relative to one of the SMBH. We classified stars into three groups: normal stars (NORM, solid lines) including MS star and WD, giant branch stars (GB, dotted lines) including RGs and AGBs and compact stars (CPT, dashed lines) including NSs and BHs. Different colors in the figure denote different mass fractions of corresponding component. Two vertical gray lines divide the evolution into three phases. The criteria for dividing three phases are based on the fluctuations of the TDE rate. Other criteria, for instance, a criteria based on dynamical evolution, will only slightly change the position and period of phase II, and will not lead to significantly different results. More details can be found in Paper II. Since NORM stars correspond to the largest fraction, the Lagrangian radii of NORM stars are almost the same as the radii of all the stars. In the left panel, CPT stars with heavier mass show significant mass segregation in the central region, especially before the formation of SMBHB. Due to the heating of SMBHB, all the Lagrangian radii in the central region expand after two SMBHs form a bound system. Especially in the phase II, a large fraction of CPT stars in the central region are kicked out or swallowed by the SMBHs, which result in a significant jump of the Lagrangian radii inside the influence radius. This result is consistent with \citet{gual12}, who has numerically investigated the evolution of SMBHBs in multi-component merging galaxies. According to the results in the right panel, only CPT stars at central region have significant evolution on average mass. NORM and GB stars roughly keep a constant average mass during the integration, with latter corresponding to heavier value. Similar to the Fig.~\ref{fig:LagrComp}, the average mass of CPT stars at inner region has a sharp drop in phase II. Since the drops of other two components are not significant, the sharp decline of the average mass in phase II should be mainly contributed by CPT stars.

\subsection{Tidal disruptions in multi-mass models}
\label{td}

\begin{figure}
\begin{center}
\includegraphics[width=0.8\textwidth,angle=0.]{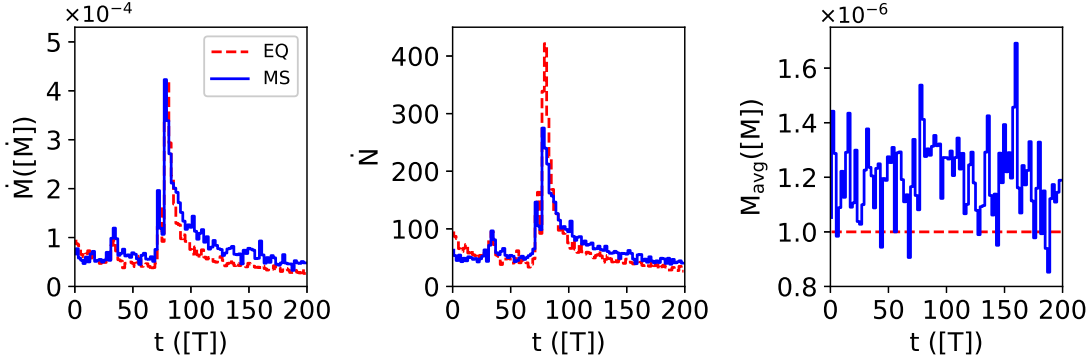}
\caption {The evolution of the total tidal disruption/swallowed rate in the multi-mass model and the equal mass model. The red dashed and blue solid lines represent the result of equal mass model and multi-mass model respectively. The left panel and middle panel are, respectively, the evolution of the mass accretion rate and the event rate. The right panel is the average mass of disrupted/swallowed stars.
\label{fig:BH-TDE3}}
\end{center}
\end{figure}

In our multi-component simulation there are 13991 TDEs/swallow events within the 200 \nbody time unit, and $\sim 94\% $ are made by MS stars. The fraction of GB, WD, NS and BH are, respectively, $\sim 2.3\% $, $\sim 2.7\% $, $\sim 0.7\% $ and $\sim 0.3\% $. Similar to the equal mass model, the tidal disruption in multi-mass models also can be divided into three phases, which correspond to before, during and after the formation of bound SMBHBs. Fig.~\ref{fig:BH-TDE3} demonstrates the total capture rate, which includes both disrupted stars and swallowed normal stars or compact stars, in the multi-mass model and the equal mass model. The red dashed and blue solid lines represent the result of equal mass model and multi-mass model respectively. The left panel and middle panel are, respectively, the evolution of the mass accretion rate and the event rate. In the simulation, we assume that all the mass of the disrupted stars will be accreted into the SMBH. That means, the evolution of the mass accretion rate and the event rate in the equal mass models are the same thing. However, it is different in multi-mass models. As a result, the peak mass accretion rates of the equal mass and multi-mass model in phase II are at the same level. While their event rates are different. Compared with the multi-mass model which has peak event rate $\dot{N}=275/[\rm T]$, the equal mass model has significantly higher peak event rate $\dot{N}=421.5/[\rm T]$. According to this result, though the disrupted/swallowed stars are less than the case in the equal mass model, the disrupted/swallowed stars in the multi-mass model are preferred at the high mass end. It has been confirmed by the right panel of Fig.~\ref{fig:BH-TDE3}, which represents the average mass of disrupted/swallowed stars. This effect is mainly due to the larger tidal radii of heavier stars, and the mass segregation may also have some contributions.

\subsubsection{Tidal disruptions of main sequence stars}
\label{tdms}

\begin{figure}
\begin{center}
\includegraphics[width=0.8\textwidth,angle=0.]{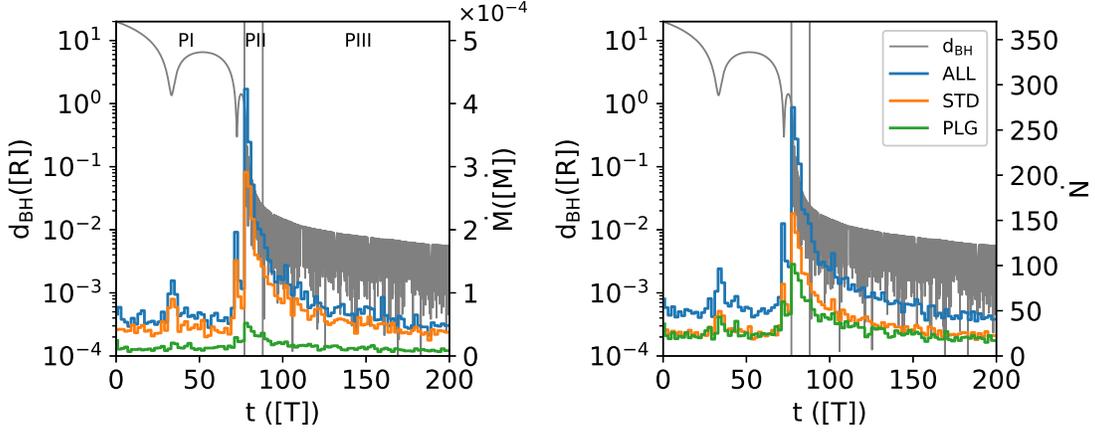}
\caption {The tidal disruption/swallow evolution of MS stars. The left panel and right panel are, respectively, the evolution of mass accretion rate and the event rate. Blue, orange and green solid lines represent the result of all of MS stars, standard TDE, and the MS stars which directly plunge into the SMBH, respectively. Two vertical gray lines divide the evolution into three phases.
\label{fig:NormTD}}
\end{center}
\end{figure}

Tidal disruption of MS stars dominate the TDEs in our models. As discussed in Section~\ref{tdschm}, MS stars with relatively low mass generally correspond to small radii. Stars with tidal radius smaller than $r_{\rm MSO}$ will be essentially swallowed when they get close enough to the SMBH. According to our simulation results with the largest particle number, there are $\sim 45\%$ MS stars, with masses range from $\sim 0.1 \msun$ to $\sim 0.4 \msun$, that have been swallowed by the SMBH. It should be noticed that all the WDs will be swallowed by the SMBH too. But they only contribute less than $3\%$ of the total disruption/swallow events. Other models with different particle numbers give similar results. In general, nearly half of captured stars will directly plunge into the SMBH without flares. Fig.~\ref{fig:NormTD} demonstrates the disruption/swallow evolution of MS stars. The orange and green lines in the figure denote stars with standard tidal disruption flare (STD) and stars with plunge orbits without flare (PLG), respectively. Blue lines and gray lines denote the evolution of all stars and the separation of two SMBHs, respectively. The entire evolution is divided into three phases by two vertical gray lines. The left y-axis represents the separation, and the right y-axis in the left and right panels are, respectively, the mass accretion rate and event number rate.

Fig.~\ref{fig:NormTD} indicates that, due to the strong perturbation and the rapid evolution of the stellar distribution around two SMBHs, both swallowed stars and normal disrupted stars have significantly enhanced rate in phase II. And there will be a rate peak every time two SMBHs get close. This result is consistent with the previous equal mass model in Paper I. From the right panel we can easily see that the normal TDEs and the plunge cases have very close event rates. However, it is obvious that the plunge stars only contribute a small fraction of the mass accretion rate in the left panel, because they are dominated by the stars at the low-mass end.

\subsubsection{Tidal disruptions of giant branch stars}
\label{gtd}

\begin{figure}
\begin{center}
\includegraphics[width=0.8\textwidth,angle=0.]{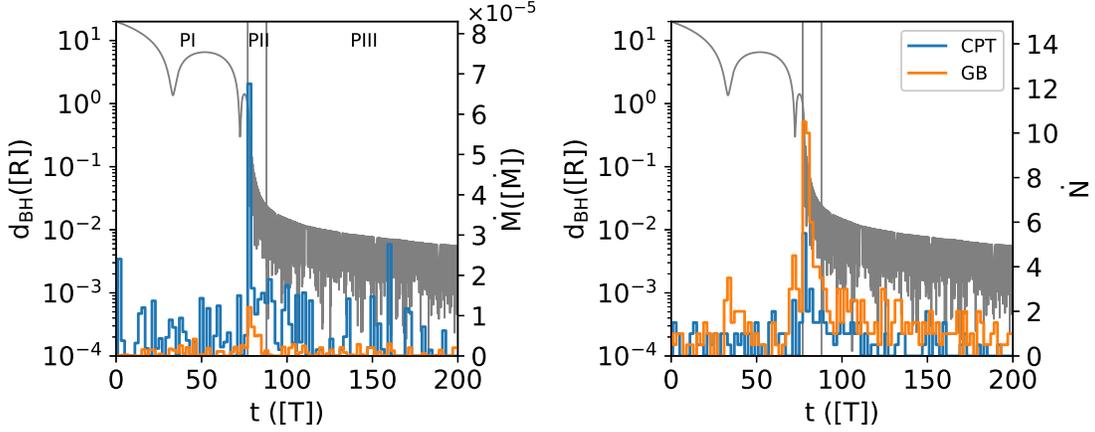}
\caption {The tidal disruption evolution of CPT stars and GB stars. The left panel and right panel are, respectively, the evolution of mass accretion rate and the event rate. Blue and orange solid lines represent the result of CPT and GB stars respectively. Two vertical gray lines divide the evolution into three phases.
\label{fig:OtherTD}}
\end{center}
\end{figure}

GB stars usually have very large radii, which means a GB star has a very large tidal radius. Usually, there is a dense core with significant mass concentrated in a GB star. During the tidal disruption, instead of being totally disrupted by the SMBH, only the envelope of the GB star will be striped away, leading to a light accretion with only a fraction of the stellar mass. As demonstrated in Fig.~\ref{fig:OtherTD}, the tidal disruption evolution of GB stars is similar to MS stars. The evolution can be divided into three phases and there is a significant enhanced rate in phase II. Here blue and orange lines represent the evolution of CPT and GB stars respectively, the rest of the legend and labels are the same as Fig.~\ref{fig:NormTD}.

\begin{figure}
\begin{center}
\includegraphics[width=0.8\textwidth,angle=0.]{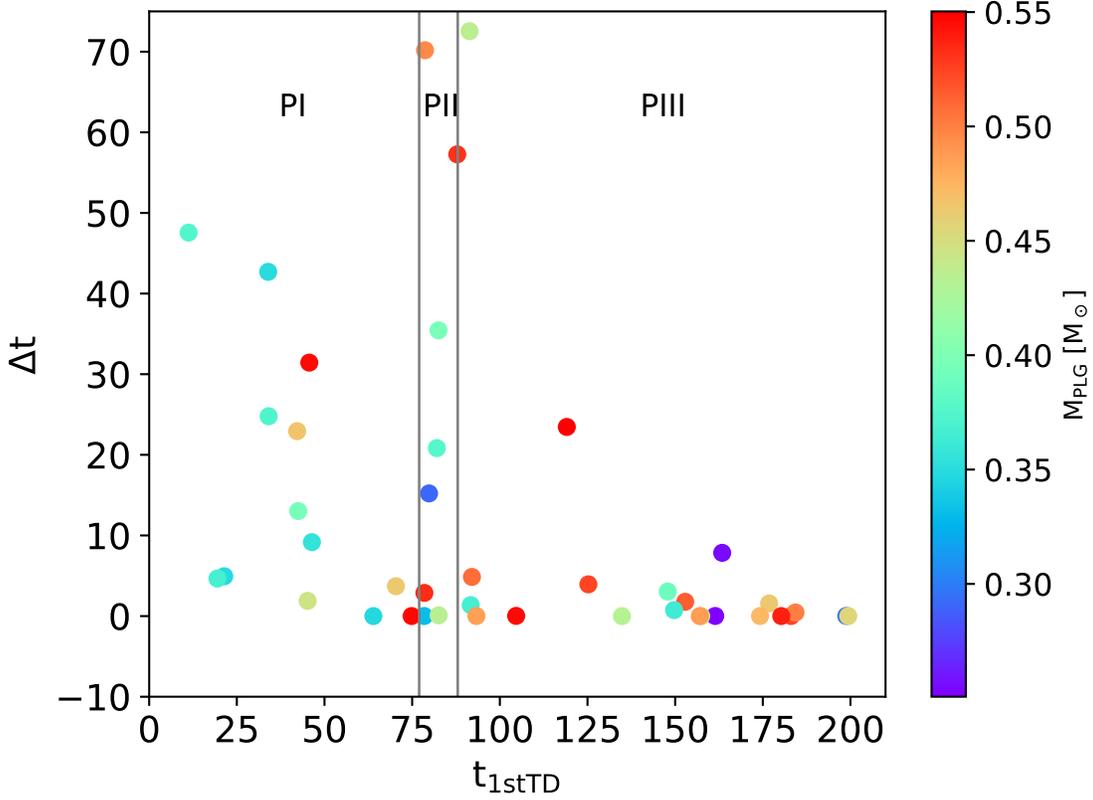}
\caption {The GB stars which have been finally swallowed by the SMBH after a tidal disruption. The x-axis and y-axis represent the time that stars have been tidally disrupted by the SMBH and the time delay between the tidal disruption and the final plunge. Two vertical gray lines divide the evolution into three phases. The color bar denotes the mass of the tidal disruption remnants which finally plunge into the SMBH, in solar mass.
\label{fig:TD207}}
\end{center}
\end{figure}

As mentioned in Section~\ref{tdschm}, after the disruption, the remnant stars usually are WDs or naked helium MS stars. Most of them will fly by the SMBH and not be swallowed. However, according to our result, there is $\sim 14\%$ remnants will finally plunge into the SMBH. Some of them will quickly plunge into the SMBH within one orbital period, and the rest may be captured by the SMBH first and finally swallowed with a time delay $\Delta t$. Fig.~\ref{fig:TD207} represents the distribution of these accreted remnant stars. The x-axis and y-axis, respectively, represent "$t_{\rm {1stTD}}$", the time that stars been tidally disrupted by the SMBH and the time delay $\Delta t$ between the first tidal disruption and the final plunge. The mass of remnants after the first tidal disruption has been denoted in different colors, with a color bar in solar mass. The size of the filled circles represents the mass of striped envelopes during GTD with the range from $2.6\times10^{-5} \msun$ to $1.8 \msun$. According to the figure, most of the plunge events of the remnants are recorded soon after the first tidal disruption. There are some exceptions which can survive for a relatively long time after the first tidal disruption. And most of such large time delay cases are in early phases. After a compact SMBHB formed in phase III, the remnants will be more easily kicked by the companion SMBH and avoid the final plunge. Since our integration terminated at $t=200$, there may be some plunge events at $t>200$ missed. In addition, we also find that $\sim 17$ stars in Fig.~\ref{fig:TD207} have been stripped off more than half of their mass during the GTD.

\subsubsection{SMBHB and compact objects}
\label{cmpt}

\begin{figure}
\begin{center}
\includegraphics[width=0.8\textwidth,angle=0.]{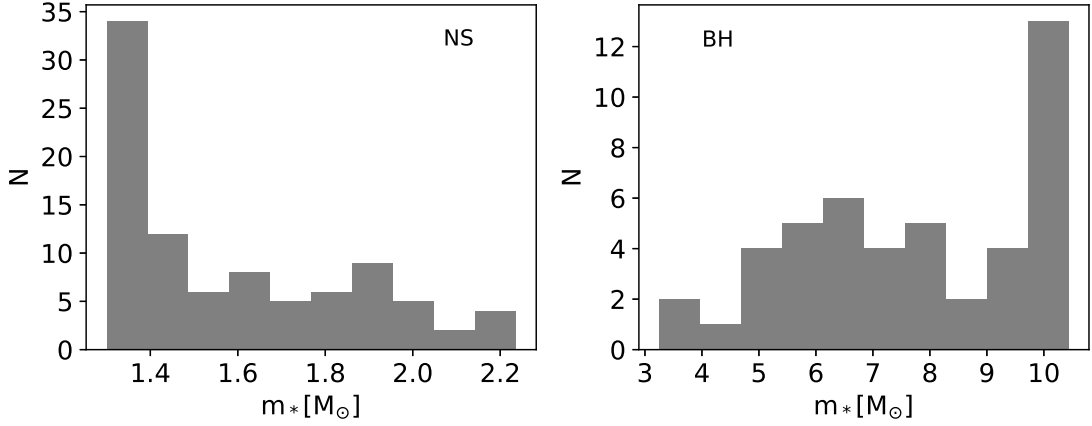}
\caption {The mass distribution of swallowed NSs and BHs. The left panel and the right panel are, respectively, the distribution of swallowed NSs and BHs. The masses of CPT stars are in solar mass and the y-axis represents the event counts.
\label{fig:BHNS}}
\end{center}
\end{figure}

Fig.~\ref{fig:OtherTD} demonstrates the swallow rate evolution of CPT stars, which is similar to GB stars. According to the heavier mass, though CPT plunge events are not as frequent as GTD, their contributions on mass accretion is more significant than GB stars. Fig.~\ref{fig:BHNS} represents the mass distribution of swallowed NSs and BHs. Most swallowed NSs concentrate in the low mass end, while the BHs are the opposite. It should be noted that, since general relativity effects are not included in all of these integrations, the orbits of compact stars very close to SMBHs are not accurate. In principle, by using post-Newtonian approximation we can integrate the orbital evolution of a two-body system with relatively high accuracy. However, such an approximation is not easy to be adopted in a triple system which we are discussing here. \citet{bone16} has achieved an improvement with corrections up to 2.5PN order. We prefer to solve this problem with similar methods in future work. This simplification will not make a significant influence on the TDR evolution. Because most of disrupted stars are tend to be concentrated on the low mass end which was relatively less affected. The rate may only be slightly underestimated, but the evolution should be the same. While for CPT stars, specially those with orbits very close to SMBHs, the absence of the PN corrections may lead to the suppressed EMRIs formation.

\section{Discussion}
\label{Diss}

\subsection{Extrapolations to galaxies}
\label{extrap}

As discussed in Section~\ref{scl}, the simulation results here can not be directly used to estimate the tidal disruption rate evolution in a galaxy. Due to the limited particle resolution, extrapolations based on the simulation results are crucial. Here we follow the same scheme as Paper I. We choose $N = 5\times10^5$ for each galaxy and the average tidal radius $\avgrt = 5\times10^{-4}$ as fiducial parameters. In order to investigate the $\rt$ and $N$ dependence, we vary $\avgrt$ with fixed fiducial $N$, and vary $N$ with fiducial $\avgrt$ individually. The $\avgrt$ varies in $[10^{-5}, 10^{-4.5}, 10^{-4}, 10^{-3.5}, 10^{-3}]$, and the $N$ varies in $[1.25\times10^5, 2.5\times10^5, 5\times10^5, 10^6]$. In every model, the average tidal disruption accretion rates and event rates in three phases are calculated individually.

\begin{figure}
\begin{center}
\includegraphics[width=0.8\textwidth,angle=0.]{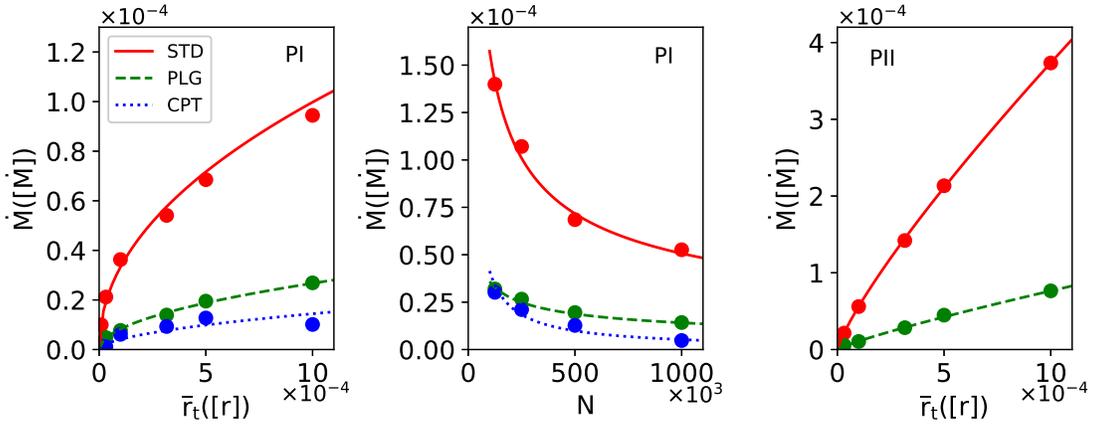}
\caption {The disruption rate extrapolations of phase I and II. The red, green and blue dots represent the tidal disruption accretion rates of STD, PLG and CPT events based on simulation results respectively. The red solid, the green dashed and the blue dotted lines represent the corresponding fitting results based on analytical estimations. The left and middle panels demonstrate the $\avgrt$ dependence and $N$ dependence in phase I, respectively. And the right panel denotes the $\avgrt$ dependence in phase II.
\label{fig:TDRFIT}}
\end{center}
\end{figure}

Fig.~\ref{fig:TDRFIT} demonstrates the corresponding simulation results of phase I and II for different stellar components. The red, green and blue dots represent the tidal disruption accretion rates of STD, PLG and CPT stars respectively. Limited by the particle resolution, even the model with largest star particles does not record enough events of GTD and CPT for statistical study in some phases. Especially in phase II, due to the short time periods, only STD and PLG stars have enough recorded events for statistical study. The left and middle panels in the figure demonstrate the $\avgrt$ dependence and $N$ dependence in phase I, respectively. And the right panel represents the $\avgrt$ dependence in phase II.

Based on the discussions in Paper I and II, the tidal disruption in phase I is dominated by two-body relaxation, and the accretion rate can be well estimated by
\beq
\dot{M} \propto \left(\frac{N}{\ln\Lambda}\right)^\alpha\avgrt^\beta,
\label{eq:MdotPI}
\eeq
where $\alpha$ and $\beta$ can be fitted through numerical simulation results, and $\ln \Lambda$ is the Coulomb logarithm and can be estimated by
\beq
\ln \Lambda \thickapprox \ln \left(\frac{\Mbh}{2\overline{m}_*}\right),
\label{eq:CL}
\eeq
where $\overline{m}_*$ is the average mass of stars \citep{pret04}.

However, unlike the equal stellar mass models, the mass accretion rates in multi-mass models can not intuitively reflect the event rate. As a rough estimation, we can simply assume $\dot{N} = \dot{M} / \overline{m}_*$. The TDE rate in phase I can be estimated by
\beq
\dot{N} \propto N\left(\frac{\ln\Lambda}{N}\right)^\zeta\avgrt^\xi,
\label{eq:NdotPI}
\eeq
where $\zeta$ and $\xi$ can be fitted through numerical simulation results.

The tidal disruption loss cone refilling in phase II is almost full, which means the tidal disruption mass accretion rate weakly depends on $N$. Therefore the rate can be approximated as a power law of $\avgrt$. But the event rate $\dot{N}$ should be divided by $\overline{m}_*$, or equivalent to multiplying by $N$. The situation is very complicated in phase III. Two-body relaxation, perturbation of companion SMBH and the triaxial stellar distributions have significant contributions to the loss cone refilling. It is very difficult to analytically estimate the disruption rate in phase III. Therefore the extrapolation in phase III is not reliable.

With above analytical models, and assuming that different stellar components roughly follow the same relations, we can make rough extrapolations for different stellar components in phase I and II.

\subsubsection{Phase I}
\label{extrapPI}
Following the similar scheme in Paper I, combining with simulation results, the tidal disruption mass accretion rate of STD stars in phase I can be estimated as

\beq
%\dot{M} \sim 1.48\times \left(\frac{N}{\ln\Lambda} \right)^{-0.5703} \avgrt^{0.4769}.
\dot{M}_{\rm STD} \sim 1.48\times \left(\frac{N}{\ln\Lambda} \right)^{-0.57} \avgrt^{0.48},
\label{eq:fitPIstd}
\eeq
which can be extrapolated as
\beq
\dot{M}_{\rm STD} \sim 2.7\times10^{-2} \left(\frac{N}{\ln\Lambda} \right)^{-0.57} \left(\frac{r_{1/2}}{1\kpc}\right)^{-1.98} \left(\frac{\avgrt}{10^{-6}\rm{pc}}\right)^{0.48} \acc.
\label{eq:expMPIstd}
\eeq

The event rate is
\beq
%\dot{N} \sim 0.8414 N \left(\frac{N}{\ln\Lambda}\right)^{-0.5703} \avgrt^{0.4769}.
\dot{N}_{\rm STD} \sim 0.84 N \left(\frac{N}{\ln\Lambda}\right)^{-0.57} \avgrt^{0.48},
\label{eq:NfitPIstd}
\eeq
which can be extrapolated as
\beq
\dot{N}_{\rm STD} \sim 1.5\times10^{-2} \frac{\msun}{\avgm} \left(\frac{N}{\ln\Lambda} \right)^{-0.57} \left(\frac{r_{1/2}}{1\kpc}\right)^{-1.98} \left(\frac{\avgrt}{10^{-6}\rm{pc}}\right)^{0.48} \rm yr^{-1}.
\label{eq:expNPIstd}
\eeq

For our fiducial model with $M=10^9\msun$, $\mbh = 10^7\msun$, $r_{1/2}=1\kpc$, and according to our simulation results with $\avgm = 0.43\msun$ corresponding to stellar radius $r_* = 0.40\rsun$, we can derive the mass accretion rate of the disrupted stars should be $\sim 1.0\times 10^{-6}\acc$, and the event rate is $\sim 1.3\times 10^{-6} \rm{yr^{-1}}$.

Similarly, the mass accretion rate of PLG stars can be estimated as
\beq
%\dot{M} \sim 0.1609\times \left(\frac{N}{\ln\Lambda} \right)^{-0.4693} \avgrt^{0.5082},
\dot{M}_{\rm PLG} \sim 0.16\times \left(\frac{N}{\ln\Lambda} \right)^{-0.47} \avgrt^{0.51},
\label{eq:fitPIplg}
\eeq
with corresponding event rate
\beq
%\dot{N} \sim 0.3201 N \left(\frac{N}{\ln\Lambda} \right)^{-0.4693} \avgrt^{0.5082}.
\dot{N}_{\rm PLG} \sim 0.32 N \left(\frac{N}{\ln\Lambda} \right)^{-0.47} \avgrt^{0.51}.
\label{eq:NfitPIplg}
\eeq

The extrapolation of the mass accretion rate and the event rate are
\beq
\dot{M}_{\rm PLG} \sim 1.6\times10^{-3} \left(\frac{N}{\ln\Lambda} \right)^{-0.47} \left(\frac{r_{1/2}}{1\kpc}\right)^{-2.01} \left(\frac{\avgrt}{10^{-6}\rm{pc}}\right)^{0.51} \acc,
\label{eq:expMPIplg}
\eeq
\beq
\dot{N}_{\rm PLG} \sim 3.2\times10^{-3}\frac{\msun}{\avgm} \left(\frac{N}{\ln\Lambda} \right)^{-0.47} \left(\frac{r_{1/2}}{1\kpc}\right)^{-2.01} \left(\frac{\avgrt}{10^{-6}\rm{pc}}\right)^{0.51} \rm yr^{-1}.
\label{eq:expNPIplg}
\eeq
In our fiducial model, the mass accretion rate should be $\sim 4.0\times 10^{-7}\acc$, and the event rate is $\sim 1.8\times 10^{-6} \rm{yr^{-1}}$.

GTD and CPT events are not popular in all simulations. GTD events are quite rare. However, we can still find some CPT stars swallowed by SMBHs in phase I. Although the data is not very good for statistical study, we can still try some extrapolations, which gives the mass accretion rate
\beq
%\dot{M} \sim 62.74\times \left(\frac{N}{\ln\Lambda} \right)^{-1.04} \avgrt^{0.5469},
\dot{M}_{\rm CPT} \sim 62.74\times \left(\frac{N}{\ln\Lambda} \right)^{-1.04} \avgrt^{0.55},
\label{eq:fitPIcpt}
\eeq
and the event rate
\beq
%\dot{N} \sim 7.178 N \left(\frac{N}{\ln\Lambda} \right)^{-1.04} \avgrt^{0.5469}.
\dot{N}_{\rm CPT} \sim 7.18 N \left(\frac{N}{\ln\Lambda} \right)^{-1.04} \avgrt^{0.55}.
\label{eq:NfitPIcpt}
\eeq

The extrapolation relations of the mass accretion rate and the event rate are
\beq
\dot{M}_{\rm CPT} \sim 0.29 \left(\frac{N}{\ln\Lambda} \right)^{-1.04} \left(\frac{r_{1/2}}{1\kpc}\right)^{-2.05} \left(\frac{\avgrt}{10^{-6}\rm{pc}}\right)^{0.55} \acc,
\label{eq:expMPIcpt}
\eeq
\beq
\dot{N}_{\rm CPT} \sim 3.3\times10^{-2}\frac{\msun}{\avgm} \left(\frac{N}{\ln\Lambda} \right)^{-1.04} \left(\frac{r_{1/2}}{1\kpc}\right)^{-2.05} \left(\frac{\avgrt}{10^{-6}\rm{pc}}\right)^{0.55} \rm yr^{-1}.
\label{eq:expNPIcpt}
\eeq
And the corresponding accretion rate and the event rate of the fiducial model are, respectively, $\dot{M} \sim 1.6\times 10^{-9}\acc$ and $\dot{N} \sim 4.4\times 10^{-10} \rm{yr^{-1}}$. It should be noticed that our results here are purely Newtonian. Since the post-Newtonian approach is not included in our model, the results here can only be considered as a crude approximation.

\begin{figure}
\begin{center}
\includegraphics[width=0.8\textwidth,angle=0.]{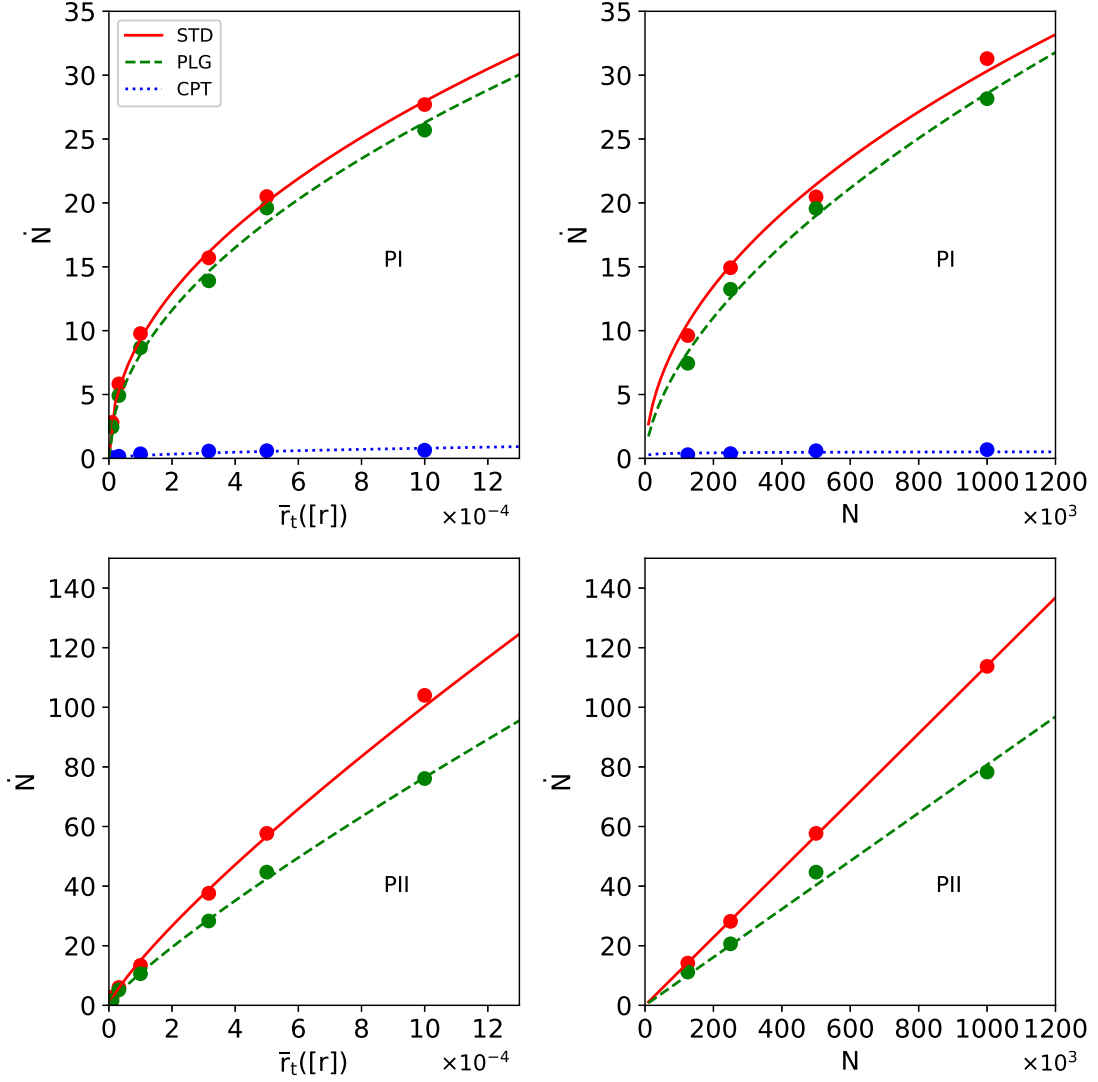}
\caption {The disruption event rate extrapolations of phase I and II. The legends are similar to the Fig.~\ref{fig:TDRFIT}. The first and the second rows demonstrate the results of phase I and phase II, respectively. The left and right panels in each row demonstrate the $\avgrt$ dependence and $N$ dependence, respectively.
\label{fig:NTDRFIT}}
\end{center}
\end{figure}

Fig.~\ref{fig:TDRFIT} and ~\ref{fig:NTDRFIT} demonstrate the extrapolation fitting results of the mass accretion rate and event rate respectively. Red, green and blue dots in two figures represent simulation results of STD, PLG and CPT stars respectively. Red solid, green dashed and blue dotted lines are, respectively, fitting results of STD, PLG and CPT stars based on the derived extrapolation formulas in Section~\ref{extrapPI} and \ref{extrapPII}. It should be aware that, due to very limited event records, the extrapolation fitting results of swallowed CPT stars in PI are not as good as STD or PLG stars. Compare the two figures, the STD and PLG have similar event rates in phase I. But the STD stars correspond to significantly higher mass accretion rates. Our extrapolation results of the fiducial model also indicate that more than half events are contributed by PLG stars, which may not induce any observational effects. But more than half accreted mass is contributed by STD stars.

\subsubsection{Phase II}
\label{extrapPII}
Both analytical estimations and our numerical simulations indicate that the tidal disruption rate in phase II does not significantly depend on $N$. The accretion rate of STD stars can be estimated by
\beq
%\dot{M} \sim 0.1107\avgrt^{0.8239},
\dot{M}_{\rm STD} \sim 0.11\avgrt^{0.82},
\label{eq:fitPIIstd}
\eeq
with extrapolation relation
\beq
\dot{M}_{\rm STD} \sim 2.2\times10^{-6} \left(\frac{r_{1/2}}{1\kpc}\right)^{-2.32} \left(\frac{\avgrt}{10^{-6}\rm{pc}}\right)^{0.82} \acc,
\label{eq:expMPIIstd}
\eeq
which corresponds to $\dot{M} \sim 5.0\times 10^{-6}\acc$ for our fiducial model.

The event rate is
\beq
% \dot{N} \sim 4.6935\times10^{-2}N\avgrt^{0.7942}.
%\dot{N} \sim 5.9666\times10^{-2}N\avgrt^{0.8239}.
\dot{N}_{\rm STD} \sim 5.97\times10^{-2}N\avgrt^{0.82},
\label{eq:NfitPIIstd}
\eeq
and the extrapolation can be wrote as
\beq
\dot{N}_{\rm STD} \sim 1.2\times10^{-6}\frac{\msun}{\avgm} \left(\frac{r_{1/2}}{1\kpc}\right)^{-2.32} \left(\frac{\avgrt}{10^{-6}\rm{pc}}\right)^{0.82} \rm yr^{-1},
\label{eq:expNPIIstd}
\eeq
which corresponds to $\dot{N} \sim 6.3\times 10^{-6} \rm{yr^{-1}}$ for the fiducial model.

Similarly, the mass accretion rate of PLG stars can be estimated by
\beq
%\dot{M} \sim 2.689\times 10^{-2}\avgrt^{0.8488},
\dot{M}_{\rm PLG} \sim 2.69\times 10^{-2}\avgrt^{0.85},
\label{eq:fitPIIplg}
\eeq
with extrapolation relation
\beq
\dot{M}_{\rm PLG} \sim 3.3\times10^{-7} \left(\frac{r_{1/2}}{1\kpc}\right)^{-2.35} \left(\frac{\avgrt}{10^{-6}\rm{pc}}\right)^{0.85} \acc.
\label{eq:expMPIIplg}
\eeq
For our fiducial model, that corresponds to $\dot{M} \sim 7.6\times 10^{-7}\acc$.

The event rate of PLG stars is
\beq
%\dot{N} \sim 4.4959\times10^{-2}N\avgrt^{0.8252}.
%\dot{N} \sim 5.3557\times10^{-2}N\avgrt^{0.8488}.
\dot{N}_{\rm PLG} \sim 5.36\times10^{-2}N\avgrt^{0.85}.
\label{eq:NfitPIIplg}
\eeq
and the extrapolation can be wrote as
\beq
\dot{N}_{\rm PLG} \sim 6.5\times10^{-7}\frac{\msun}{\avgm} \left(\frac{r_{1/2}}{1\kpc}\right)^{-2.35} \left(\frac{\avgrt}{10^{-6}\rm{pc}}\right)^{0.85} \rm yr^{-1},
\label{eq:expNPIIplg}
\eeq
which corresponding to $\dot{N} \sim 3.5\times 10^{-6} \rm{yr^{-1}}$ for the fiducial model. Obviously, both the mass accretion rate and the event rate in phase II have significant increases.  The rate of STD has increased several times compared with PI, which is consistent with the result in Paper I and II.

Since the period of phase II in all of our models is around 20 \nbody time units, there are not many recorded events contributed by CPT and GTD stars. The extrapolation can not be well managed for them.

Table.~\ref{tab:rate} summarizes the extrapolation results of averaged mass accretion rates and event rates in phase I and II, for disrupted/swallowed STD, PLG and CPT stars respectively. Since the swallowed CPT stars in phase II are relatively rare, we do not estimate their rates in the table.

\begin{deluxetable}{ccccccc}
    \tablewidth{0pt}
    \tabletypesize{\scriptsize}
    \tablecaption{Averaged mass accretion rates and event rates in different phases. \label{tab:rate}}
    \tablehead{
    \\
    \colhead{Stage} &
    \colhead{$\dot{M}_{\rm STD}$} &
    \colhead{$\dot{N}_{\rm STD}$} &
    \colhead{$\dot{M}_{\rm PLG}$} &
    \colhead{$\dot{N}_{\rm PLG}$} &
    \colhead{$\dot{M}_{\rm CPT}$} &
    \colhead{$\dot{N}_{\rm CPT}$} \\
    \colhead{} &
    \colhead{$\times 10^{-6}\rm M_\odot/yr$} &
    \colhead{$\times 10^{-6}\rm{/yr}$} &
    \colhead{$\times 10^{-6}\rm M_\odot/yr$} &
    \colhead{$\times 10^{-6}\rm{/yr}$} &
    \colhead{$\times 10^{-9}\rm M_\odot/yr$} &
    \colhead{$\times 10^{-9}\rm{/yr}$} \\
    \colhead{(1)} &
    \colhead{(2)} &
    \colhead{(3)} &
    \colhead{(4)} &
    \colhead{(5)} &
    \colhead{(6)} &
    \colhead{(7)}
    }
    \startdata
PI & 1.0 & 1.3 & 0.4 & 1.8 & 1.6 & 0.4  \\
PII & 5.0 & 6.3 & 0.8 & 3.5 & - & - \\
    \enddata

\tablecomments{Col.(1): Stage of the evolution. Col.(2): Mass accretion rate of disrupted STD stars. Col.(3): Event rate of disrupted STD stars. Col.(4): Mass accretion rate of swallowed PLG stars. Col.(5): Event rate of swallowed PLG stars. Col.(6): Mass accretion rate of swallowed CPT stars. Col.(7): Event rate of swallowed CPT stars.}
\end{deluxetable}

\subsection{Delectability by space based GW instruments}
\label{gw}

As strong GW sources, SMBHBs could be directly detected by PTAs or space born detectors such as LISA, TaiJi and TianQin in the future. There are many discussions on this topic, which is out of the scope of this paper. Here we want to discuss EMRIs, another kind of typical GW source produced by SMBHs which could be detected by LISA/TaiJi/TianQin. Such events could also be prompted around a SMBHB. However, a typical EMRI should spiral the SMBH for many orbits. It needs accurate integrations with general relativity effects considered, which is not included in our integrations. Therefore, it is not reliable to estimate the rate of EMRIs based on our models. Another interesting event are EMRBs, which are GW bursts from stellar objects passing by a SMBH with very small pericenter distances. It has been considered as the precursor to EMRIs, because many EMRBs will lose their energy and angular momentum through GW radiation and finally evolve into EMRIs \citep{rub06}. \citet{hop07} demonstrates that the burst rates for stellar BHs and MSs/WDs are, respectively, $1 \yr^{-1}$ and $0.1 \yr^{-1}$ in the Milky Way. If extragalactic sources could be included, a detector like LISA could manage the detection out to $\sim 100 \Mpc$ for a $10 \msun$ BH, with event rate of $\sim 0.2 \yr^{-1}$ \citep{ber13b, ber13c}. Recently, \citet{han20} calculates the event rates of very extreme mass ratio bursts with a mass ratio about $10^{-8}$. They especially considered the contribution of plunge stellar objects such as brown dwarfs with unbound orbits. Their estimation indicates that, for small stellar objects with mass $\sim 0.1\msun$, the space based facilities could detect the bursts inside $10 \Mpc$, with corresponding event rate $4-8 \yr^{-1}$.

In our model, as discussed in Section~\ref{extrap}, the rates of compact stars such as NSs and BHs getting swallowed by SMBHs are quite low. However, there are many low mass main sequence stars that plunge into SMBHs. According to the result of our largest simulation, the average mass of swallowed stars is $\sim 0.2 \msun$, which have similar mass as \citet{han20} discussed. In principal, the signal-to-noise ratio (SNR) of such kind of EMRBs is proportional to $\mstar R^{-1}\Mbh^{2/3}$, with $R$ is the distance from the source to observers \citep{ber13b}. With carefully numerical estimations, \citet{han20} finds that the SNRs of plunging events with $0.1 \msun$ stars in the Galactic Center can be up to ten thousands for LISA. If we consider a $0.2 \msun$ star swallowed by an $10^7 \msun$ SMBH, the SNR can be still as large as $\sim 8$ at $\sim 50\Mpc$. However, according to estimations in Section~\ref{extrap}, the event rates of the PLG are only around $10^{-6} \yr^{-1}$ in phase I and II.  Since there are not too many galaxies inside $\sim 50\Mpc$, it is unlikely to detect such kind of EMRBs with space borne GW detectors in the near future.

\section{Summary}
\label{sum}

We investigated full and partial tidal disruption of stars and direct plunges into supermassive black holes (MSBH) in nuclear star clusters during and after a galaxy merger. For that a full direct \nbody simulation has been used with stars obtained from a realistic stellar mass distribution, evolved for $\sim 1 \Gyr$ (to account for the age of the galaxies before the merger), surrounding two SMBH situated in the centres of the two merging galaxies.

With the stellar evolution included, there are different stellar components, from low mass main sequence stars to white dwarfs, neutron stars and stellar mass black holes. Different stellar objects, according to their properties like mass and radius, have different fates after the close encounter with central SMBHs. Compared to the equal mass model, the multi-mass model with similar parameters tends to have similar or even slightly higher mass accretion rate. However, although their event rates in phase I and III (before and after the galaxy merger) are similar, the multi-mass model has lower event rate compared to the equal mass model in phase II (the short time while the merger is dynamically ongoing), which indicates that the disrupted/swallowed stars in multi-mass model prefer high mass end.

In the multi-mass model, if a main sequence star is heavy enough, its close encounter with a SMBH may lead to a standard TDE. Otherwise a light main sequence star may correspond to a tiny tidal radius which makes it inside the marginally bound orbits. That will result in a plunge event. During the galaxy merger, due to the perturbation of the companion SMBH and stars around it, both STD and PLG event rates will be enhanced in phase II. STD and PLG have similar event rates. But STD events correspond to significantly higher mass accretion rate, because the PLG events are preferring to low mass stars. Since PLG events may not have any observational signatures, probably nearly half of the SMBH tidal capture events are invisible.

Post sequence stars, such as RGs or AGB stars, could be partially disrupted. There might be a core that survived after a GTD which stripped away the envelope. The remnant could escape or plunge into the SMBH. Our largest numerical simulation gets $\sim 10\%$ GTD stars finally plunge into SMBHs after their disruption. Some of them can have bound orbits around the SMBH and survive for many orbits.

CPT stars show significant mass segregation in the central region during phase I. Due to the heating of the newly formed SMBHB, the Lagrangian radii in the central region expand quickly during phase II. Since our integrations are totally Newtonian, the orbits of compact stars close to the SMBH are not accurate. According to our limited results, the rate of compact stars getting swallowed by SMBHs in phase II could be significantly enhanced. And most swallowed NSs concentrate on the low mass end, while BHs are the opposite.

\begin{acknowledgments}
We are grateful to the support of the National Natural Science Foundation of China (NSFC11988101,NSFC11303039), the Key International Partnership Program of the Chinese Academy of Sciences (CAS) (No.114A11KYSB20170015), and the Strategic Priority Research Program (Pilot B) Multiwavelength gravitational wave universe of CAS (No.XDB23040100). We (SL, PB, RS) acknowledge support by CAS through the Silk Road Project at National Astronomical Observatories (NAOC) of China, and the support by Key Laboratory of Computational Astrophysics. The computations have been done on the Laohu supercomputer at the Center of Information and Computing at NAOC, CAS, funded by the Ministry of Finance of People's Republic of China under the grant $ZDYZ2008-2$. LS and RS acknowledge the support of Yunnan Academician Workstation of Wang Jingxiu (No. 202005AF150025). LS acknowledges support from the K.C.Wong Education Foundation. PB acknowledges the special support by the CAS President's International Fellowship for Visiting Scientists (PIFI) program during his stay in NAOC, CAS. XC acknowledges the support of the National Natural Science Foundation of China (No. 11873022).
The work of PB was supported by the Volkswagen Foundation under the special stipend No.~9B870 and the grant No.~97778.
PB acknowledge the support within the grant No.~AP14870501
of the Science Committee of the Ministry of Science and
Higher Education of Kazakhstan.
The work of PB was supported under the special program of the NRF of Ukraine Leading and Young Scientists Research Support - ”Astrophysical Relativistic Galactic Objects (ARGO): life cycle of active nucleus”, No.~2020.02/0346.
PB thanks the support from the ACIISI, Consejer\'{i}a de Econom\'{i}a, Conocimiento y Empleo del Gobierno de Canarias and the European Regional Development Fund (ERDF)
under grant with reference PROID2021010044.
\end{acknowledgments}


\begin{thebibliography}{dummy}

\bibitem[Amaro-Seoane(2018)]{pau18} Amaro-Seoane, P.\ 2018, Living Reviews in Relativity, 21, 4. doi:10.1007/s41114-018-0013-8

\bibitem[Arcavi et al.(2014)]{arca14} Arcavi, I., Gal-Yam, A., Sullivan, M., et al.\ 2014, \apj, 793, 38. doi:10.1088/0004-637X/793/1/38

\bibitem[Arzoumanian et al.(2016)]{arzo16} Arzoumanian, Z., Brazier, A., Burke-Spolaor, S., et al.\ 2016, \apj, 821, 13. doi:10.3847/0004-637X/821/1/13

\bibitem[Arzoumanian et al.(2020)]{arzo20} Arzoumanian, Z., Baker, P.~T., Blumer, H., et al.\ 2020, \apjl, 905, L34. doi:10.3847/2041-8213/abd401

\bibitem[Babak et al.(2016)]{baba16} Babak, S., Petiteau, A., Sesana, A., et al.\ 2016, \mnras, 455, 1665. doi:10.1093/mnras/stv2092

\bibitem[Baumgardt et al.(2004)]{bau04} Baumgardt, H., Makino, J., \& Ebisuzaki, T.\ 2004, \apj, 613, 1143

\bibitem[Begelman et al.(1980)]{bege80} Begelman, M.~C., Blandford, R.~D., \& Rees, M.~J.\ 1980, \nat, 287, 307

\bibitem[Belczynski et al.(2002)]{belc02} Belczynski, K., Kalogera, V., \& Bulik, T.\ 2002, \apj, 572, 407. doi:10.1086/340304

\bibitem[Berczik et al.(2005)]{ber05} Berczik, P., Merritt, D., \& Spurzem, R.\ 2005, \apj, 633, 680. doi:10.1086/491598

\bibitem[Berczik et al.(2006)]{ber06} Berczik, P., Merritt, D., Spurzem, R., \& Bischof, H.-P.\ 2006, \apjl, 642, L21. doi:10.1086/504426

\bibitem[Berry \& Gair(2013b)]{ber13b} Berry, C.~P.~L. \& Gair, J.~R.\ 2013b, \mnras, 433, 3572. doi:10.1093/mnras/stt990

\bibitem[Berry \& Gair(2013c)]{ber13c} Berry, C.~P.~L. \& Gair, J.~R.\ 2013c, \mnras, 435, 3521. doi:10.1093/mnras/stt1543

\bibitem[Bonetti et al.(2016)]{bone16} Bonetti, M., Haardt, F., Sesana, A., et al.\ 2016, \mnras, 461, 4419. doi:10.1093/mnras/stw1590

\bibitem[Chatterjee et al.(2003)]{chat03} Chatterjee, P., Hernquist, L., \& Loeb, A.\ 2003, \apj, 592, 32

\bibitem[Chen et al.(2009)]{chen09} Chen, X., Madau, P., Sesana, A., \& Liu, F.~K.\ 2009, \apjl, 697, L149

\bibitem[Chevalier(1989)]{chev89} Chevalier, R.~A.\ 1989, \apj, 346, 847. doi:10.1086/168066

\bibitem[Colgate(1971)]{colg71} Colgate, S.~A.\ 1971, \apj, 163, 221. doi:10.1086/150760

\bibitem[Colpi(2014)]{colp14} Colpi, M.\ 2014, \ssr, 183, 189. doi:10.1007/s11214-014-0067-1

\bibitem[Coughlin et al.(2017)]{coug17} Coughlin, E.~R., Armitage, P.~J., Nixon, C., \& Begelman, M.~C.\ 2017, \mnras, 465, 3840. doi:10.1093/mnras/stw2913

\bibitem[Cutler et al.(1994)]{cutl94} Cutler, C., Kennefick, D., \& Poisson, E.\ 1994, \prd, 50, 3816. doi:10.1103/PhysRevD.50.3816

\bibitem[Dehnen(1993)]{deh93} Dehnen, W.\ 1993, \mnras, 265, 250

\bibitem[Evans \& Kochanek(1989)]{evan89} Evans, C.~R., \& Kochanek, C.~S.\ 1989, \apjl, 346, L13

\bibitem[Ferrarese \& Merritt(2000)]{ferr00} Ferrarese, L., \& Merritt, D.\ 2000, \apjl, 539, L9

\bibitem[Foster \& Backer(1990)]{fost90} Foster, R.~S., \& Backer, D.~C.\ 1990, \apj, 361, 300. doi:10.1086/169195

\bibitem[Frank \& Rees(1976)]{fra76} Frank, J., \& Rees, M.~J.\ 1976, \mnras, 176, 633. doi:10.1093/mnras/176.3.633

\bibitem[Gair et al.(2005)]{gair05} Gair, J.~R., Kennefick, D.~J., \& Larson, S.~L.\ 2005, \prd, 72, 084009. doi:10.1103/PhysRevD.72.084009

\bibitem[Gebhardt et al.(2000)]{gebh00} Gebhardt, K., Bender, R., Bower, G., et al.\ 2000, \apjl, 539, L13

\bibitem[Gezari(2021)]{geza21} Gezari, S.\ 2021, \araa, 59, 21, doi:10.1146/annurev-astro-111720-030029

\bibitem[Gould \& Rix(2000)]{goul00} Gould, A., \& Rix, H.-W.\ 2000, \apjl, 532, L29. doi:10.1086/312562

\bibitem[Gualandris \& Merritt(2008)]{gual08} Gualandris, A., \& Merritt, D.\ 2008, \apj, 678, 780. doi:10.1086/586877

\bibitem[Gualandris \& Merritt(2012)]{gual12} Gualandris, A. \& Merritt, D.\ 2012, \apj, 744, 74. doi:10.1088/0004-637X/744/1/74

\bibitem[Guillochon \& Ramirez-Ruiz(2013)]{guil13} Guillochon, J., \& Ramirez-Ruiz, E.\ 2013, \apj, 767, 25

\bibitem[Han et al.(2020)]{han20} Han, W.-B., Zhong, X.-Y., Chen, X., et al.\ 2020, \mnras, 498, L61. doi:10.1093/mnrasl/slaa115

\bibitem[Harfst et al.(2007)]{harf07} Harfst, S., Gualandris, A., Merritt, D., Spurzem, R., Portegies Zwart, S., \& Berczik, P.\ 2007, New Astronomy, 12, 357. doi:10.1016/j.newast.2006.11.003

\bibitem[Hills(1975)]{hil75} Hills, J.~G.\ 1975, \nat, 254, 295. doi:10.1038/254295a0

\bibitem[Hobbs et al.(2005)]{hobb05} Hobbs, G., Lorimer, D.~R., Lyne, A.~G., et al.\ 2005, \mnras, 360, 974. doi:10.1111/j.1365-2966.2005.09087.x

\bibitem[Hopman et al.(2007)]{hop07} Hopman, C., Freitag, M., \& Larson, S.~L.\ 2007, \mnras, 378, 129. doi:10.1111/j.1365-2966.2007.11758.x

\bibitem[Huang et al.(2021)]{huan21} Huang, S., Hu, S., Yin, H., et al.\ 2021, \apj, 920, 12. doi:10.3847/1538-4357/ac0eff

\bibitem[Hurley et al.(2000)]{hurl00} Hurley, J.~R., Pols, O.~R., \& Tout, C.~A.\ 2000, \mnras, 315, 543. doi:10.1046/j.1365-8711.2000.03426.x

\bibitem[Ivanov et al.(2005)]{ivan05} Ivanov, P.~B., Polnarev, A.~G., \& Saha, P.\ 2005, \mnras, 358, 1361. doi:10.1111/j.1365-2966.2005.08843.x

\bibitem[Khan et al.(2018)]{khan18} Khan, F.~M., Berczik, P., \& Just, A.\ 2018, \aap, 615, A71. doi:10.1051/0004-6361/201730489

\bibitem[Khan et al.(2011)]{khan11} Khan, F.~M., Just, A., \& Merritt, D.\ 2011, \apj, 732, 89. doi:10.1088/0004-637X/732/2/89

\bibitem[Komossa(2015)]{komo15} Komossa, S.\ 2015, Journal of High Energy Astrophysics, 7, 148. doi:10.1016/j.jheap.2015.04.006

\bibitem[Komossa \& Bade(1999)]{komo99} Komossa, S., \& Bade, N.\ 1999, \aap, 343, 775.

\bibitem[Komossa et al.(2003)]{komo03} Komossa, S., Burwitz, V., Hasinger, G., et al.\ 2003, \apjl, 582, L15. doi:10.1086/346145

\bibitem[Komossa et al.(2020)]{komo20} Komossa, S., Grupe, D., Parker, M.~L., et al.\ 2020, \mnras, 498, L35. doi:10.1093/mnrasl/slaa125

\bibitem[Kormendy \& Ho(2013)]{korm13} Kormendy, J., \& Ho, L.~C.\ 2013, \araa, 51, 511

\bibitem[Kroupa(2001)]{krou01} Kroupa, P.\ 2001, \mnras, 322, 231. doi:10.1046/j.1365-8711.2001.04022.x

\bibitem[Li et al.(2017)]{li17} Li, S., Liu, F.~K., Berczik, P., \& Spurzem, R.\ 2017, \apj, 834, 195

\bibitem[Li et al.(2019)]{li19} Li, S., Berczik, P., Chen, X., Liu, F.~K., Spurzem, R., \& Qiu, Y.\ 2019, \apj, 883, 132

\bibitem[Lightman \& Shapiro(1977)]{lig77} Lightman, A.~P., \& Shapiro, S.~L.\ 1977, \apj, 211, 244. doi:10.1086/154925

\bibitem[Liu(2004)]{liu04} Liu, F.~K.\ 2004, \mnras, 347, 1357. doi:10.1111/j.1365-2966.2004.07325.x

\bibitem[Liu \& Chen(2013)]{liu13} Liu, F.~K., \& Chen, X.\ 2013, \apj, 767, 18. doi:10.1088/0004-637X/767/1/18

\bibitem[Liu \& Wu(2002)]{liu02} Liu, F.~K., \& Wu, X.-B.\ 2002, \aap, 388, L48. doi:10.1051/0004-6361:20020566

\bibitem[Liu et al.(2009)]{liu09} Liu, F.~K., Li, S., \& Chen, X.\ 2009, \apjl, 706, L133

\bibitem[Liu et al.(2014)]{liu14} Liu, F.~K., Li, S., \& Komossa, K. \ 2014, \apj, 786, 103

\bibitem[Luo et al.(2021)]{luo21} Luo, Z., Wang, Y., Wu, Y., et al.\ 2021, Progress of Theoretical and Experimental Physics, 2021, 05A108. doi:10.1093/ptep/ptaa083

\bibitem[MacLeod et al.(2012)]{macl12} MacLeod, M., Guillochon, J., \& Ramirez-Ruiz, E.\ 2012, \apj, 757, 134. doi:10.1088/0004-637X/757/2/134

\bibitem[Makino \& Aarseth(1992)]{maki92} Makino, J., \& Aarseth, S.~J.\ 1992, \pasj, 44, 141

\bibitem[Magorrian et al.(1998)]{mago98} Magorrian, J., Tremaine, S., Richstone, D., et al.\ 1998, \aj, 115, 2285

\bibitem[Mei et al.(2021)]{mei21} Mei, J., Bai, Y.-Z., Bao, J., et al.\ 2021, Progress of Theoretical and Experimental Physics, 2021, 05A107. doi:10.1093/ptep/ptaa114

\bibitem[Merritt \& Poon(2004)]{mer04} Merritt, D., \& Poon, M.~Y.\ 2004, \apj, 606, 788

\bibitem[Mikkola \& Valtonen(1992)]{mikk92} Mikkola, S., \& Valtonen, M.~J.\ 1992, \mnras, 259, 115

\bibitem[Milosavljevi{\'c} \& Merritt(2003)]{milo03} Milosavljevi{\'c}, M., \& Merritt, D. 2003, in American Institute of Physics Conference Series, Vol. 686, The Astrophysics of Gravitational Wave Sources, ed. J.~M. Centrella, 201--210, doi:10.1063/1.1629432

\bibitem[Panamarev et al.(2018)]{pana18} Panamarev, T., Shukirgaliyev, B., Meiron, Y., et al.\ 2018, \mnras, 476, 4224. doi:10.1093/mnras/sty459

\bibitem[Panamarev et al.(2019)]{pana19} Panamarev, T., Just, A., Spurzem, R., et al.\ 2019, \mnras, 484, 3279. doi:10.1093/mnras/stz208

\bibitem[Preto et al.(2004)]{pret04} Preto, M., Merritt, D., \& Spurzem, R.\ 2004, \apjl, 613, L109

\bibitem[Preto et al.(2011)]{pret11} Preto, M., Berentzen, I., Berczik, P., \& Spurzem, R.\ 2011, \apjl, 732, L26. doi:10.1088/2041-8205/732/2/L26

\bibitem[Quinlan(1996)]{quin96} Quinlan, G.~D.\ 1996, \na, 1, 35

\bibitem[Rees(1988)]{ree88} Rees, M.~J.\ 1988, \nat, 333, 523

\bibitem[Ricarte et al.(2016)]{rica16} Ricarte, A., Natarajan, P., Dai, L., et al.\ 2016, \mnras, 458, 1712. doi:10.1093/mnras/stw355

\bibitem[Rubbo et al.(2006)]{rub06} Rubbo, L.~J., Holley-Bockelmann, K., \& Finn, L.~S.\ 2006, \apjl, 649, L25. doi:10.1086/508326

\bibitem[Saslaw et al.(1974)]{sasl74} Saslaw, W.~C., Valtonen, M.~J., \& Aarseth, S.~J.\ 1974, \apj, 190, 253

\bibitem[Shannon et al.(2015)]{shan15} Shannon, R.~M., Ravi, V., Lentati, L.~T., et al.\ 2015, Science, 349, 1522. doi:10.1126/science.aab1910

\bibitem[Shen et al.(2013)]{shen13} Shen, Y., Liu, X., Loeb, A., \& Tremaine, S.\ 2013, \apj, 775, 49. doi:10.1088/0004-637X/775/1/49

\bibitem[Shu et al.(2020)]{shu20} Shu, X., Zhang, W., Li, S., et al.\ 2020, Nature Communications, 11, 5876. doi:10.1038/s41467-020-19675-z

\bibitem[Tang et al.(2021)]{tang21} Tang, S., Silverman, J.~D., Ding, X., et al.\ 2021, \apj, 922, 83. doi:10.3847/1538-4357/ac1ff0

\bibitem[Tremaine et al.(2002)]{trem02} Tremaine, S., Gebhardt, K., Bender, R., et al.\ 2002, \apj, 574, 740

\bibitem[Verbiest et al.(2016)]{verb16} Verbiest, J.~P.~W., Lentati, L., Hobbs, G., et al.\ 2016, \mnras, 458, 1267. doi:10.1093/mnras/stw347

\bibitem[Vigneron et al.(2018)]{vign18} Vigneron, Q., Lodato, G., \& Guidarelli, A.\ 2018, \mnras, 476, 5312

\bibitem[Volonteri et al.(2003)]{volo03} Volonteri, M., Haardt, F., \& Madau, P.\ 2003, \apj, 582, 559

\bibitem[Wegg \& Nate Bode(2011)]{wegg11} Wegg, C., \& Nate Bode, J.\ 2011, \apjl, 738, L8. doi:10.1088/2041-8205/738/1/L8

\bibitem[Yu(2002)]{yu02} Yu, Q.\ 2002, \mnras, 331, 935

\bibitem[Zhang et al.(2008)]{zhan08} Zhang, W., Woosley, S.~E., \& Heger, A.\ 2008, \apj, 679, 639. doi:10.1086/526404

\bibitem[Zhong et al.(2014)]{zhong14} Zhong, S., Berczik, P., \& Spurzem, R.\ 2014, \apj, 792, 137

\bibitem[Zhong et al.(2022)]{zhong22} Zhong, S., Li, S., Berczik, P., et al.\ 2022, \apj, 933, 96


\end{thebibliography}
\end{document}